\PassOptionsToPackage{unicode}{hyperref}
\PassOptionsToPackage{hyphens}{url}
\PassOptionsToPackage{dvipsnames,svgnames,x11names}{xcolor}
\documentclass[
]{article}

\usepackage{amsmath,amssymb}
\usepackage{iftex}
\ifPDFTeX
  \usepackage[T1]{fontenc}
  \usepackage[utf8]{inputenc}
  \usepackage{textcomp} 
\else 
  \usepackage{unicode-math}
  \defaultfontfeatures{Scale=MatchLowercase}
  \defaultfontfeatures[\rmfamily]{Ligatures=TeX,Scale=1}
\fi
\usepackage{lmodern}
\ifPDFTeX\else  
\fi
\IfFileExists{upquote.sty}{\usepackage{upquote}}{}
\IfFileExists{microtype.sty}{
  \usepackage[]{microtype}
  \UseMicrotypeSet[protrusion]{basicmath} 
}{}
\makeatletter
\@ifundefined{KOMAClassName}{
  \IfFileExists{parskip.sty}{%
    \usepackage{parskip}
  }{
    \setlength{\parindent}{0pt}
    \setlength{\parskip}{6pt plus 2pt minus 1pt}}
}{
  \KOMAoptions{parskip=half}}
\makeatother
\usepackage{xcolor}
\makeatletter
\ifx\paragraph\undefined\else
  \let\oldparagraph\paragraph
  \renewcommand{\paragraph}{
    \@ifstar
      \xxxParagraphStar
      \xxxParagraphNoStar
  }
  \newcommand{\xxxParagraphStar}[1]{\oldparagraph*{#1}\mbox{}}
  \newcommand{\xxxParagraphNoStar}[1]{\oldparagraph{#1}\mbox{}}
\fi
\ifx\subparagraph\undefined\else
  \let\oldsubparagraph\subparagraph
  \renewcommand{\subparagraph}{
    \@ifstar
      \xxxSubParagraphStar
      \xxxSubParagraphNoStar
  }
  \newcommand{\xxxSubParagraphStar}[1]{\oldsubparagraph*{#1}\mbox{}}
  \newcommand{\xxxSubParagraphNoStar}[1]{\oldsubparagraph{#1}\mbox{}}
\fi
\makeatother

\usepackage{longtable,booktabs,array}
\usepackage{calc} 
\usepackage{etoolbox}
\makeatletter
\patchcmd\longtable{\par}{\if@noskipsec\mbox{}\fi\par}{}{}
\makeatother
\IfFileExists{footnotehyper.sty}{\usepackage{footnotehyper}}{\usepackage{footnote}}
\makesavenoteenv{longtable}
\usepackage{graphicx}
\makeatletter
\newsavebox\pandoc@box
\newcommand*\pandocbounded[1]{
  \sbox\pandoc@box{#1}%
  \Gscale@div\@tempa{\textheight}{\dimexpr\ht\pandoc@box+\dp\pandoc@box\relax}%
  \Gscale@div\@tempb{\linewidth}{\wd\pandoc@box}%
  \ifdim\@tempb\p@<\@tempa\p@\let\@tempa\@tempb\fi
  \ifdim\@tempa\p@<\p@\scalebox{\@tempa}{\usebox\pandoc@box}%
  \else\usebox{\pandoc@box}%
  \fi%
}
\def\fps@figure{htbp}
\makeatother
\NewDocumentCommand\citeproctext{}{}
\NewDocumentCommand\citeproc{mm}{%
  \begingroup\def\citeproctext{#2}\cite{#1}\endgroup}
\makeatletter
 \let\@cite@ofmt\@firstofone
 \def\@biblabel#1{}
 \def\@cite#1#2{{#1\if@tempswa , #2\fi}}
\makeatother
\newlength{\cslhangindent}
\newlength{\csllabelwidth}
\newenvironment{CSLReferences}[2] 
 {\begin{list}{}{%
  \setlength{\itemindent}{0pt}
  \setlength{\leftmargin}{0pt}
  \setlength{\parsep}{0pt}
  \ifodd #1
   \setlength{\leftmargin}{\cslhangindent}
   \setlength{\itemindent}{-1\cslhangindent}
  \fi
  \setlength{\itemsep}{#2\baselineskip}}}
 {\end{list}}
\usepackage{calc}

\makeatletter
\renewcommand{\maketitle}{%
  \begin{center}
  {\Large\bfseries \@title\par}\vspace{1em}
  Aleksandar Tomašević\textsuperscript{1,*}, Hudson Golino\textsuperscript{2}, and
  Alexander P. Christensen\textsuperscript{3}\par\vspace{0.5em}
  {\small
  \textsuperscript{1}Institute of Physics Belgrade, University of Belgrade\par
  \textsuperscript{2}Department of Psychology, University of Virginia\par
  \textsuperscript{3}Department of Psychology and Human Development, Peabody College, Vanderbilt University\par}
  \end{center}
  \noindent \textsuperscript{*}Corresponding author: \href{mailto:atomasevic@ipb.ac.rs}{atomasevic@ipb.ac.rs}.\par
  \vspace{1em}
}
\makeatother
\usepackage{setspace}
\usepackage{placeins}
\usepackage{float}
\usepackage{hyperref}
\hypersetup{colorlinks=true, linkcolor=blue, urlcolor=blue, citecolor=blue}
\usepackage[margin=1in]{geometry}
\usepackage{lineno}
\usepackage{amsmath}
\usepackage{booktabs}
\usepackage{pdflscape}
\makeatletter
\@ifpackageloaded{caption}{}{\usepackage{caption}}
\AtBeginDocument{%
\ifdefined\contentsname
  \renewcommand*\contentsname{Table of contents}
\else
  \newcommand\contentsname{Table of contents}
\fi
\ifdefined\listfigurename
  \renewcommand*\listfigurename{List of Figures}
\else
  \newcommand\listfigurename{List of Figures}
\fi
\ifdefined\listtablename
  \renewcommand*\listtablename{List of Tables}
\else
  \newcommand\listtablename{List of Tables}
\fi
\ifdefined\figurename
  \renewcommand*\figurename{Figure}
\else
  \newcommand\figurename{Figure}
\fi
\ifdefined\tablename
  \renewcommand*\tablename{Table}
\else
  \newcommand\tablename{Table}
\fi
}
\@ifpackageloaded{float}{}{\usepackage{float}}
\floatstyle{ruled}
\@ifundefined{c@chapter}{\newfloat{codelisting}{h}{lop}}{\newfloat{codelisting}{h}{lop}[chapter]}
\floatname{codelisting}{Listing}

\makeatother
\makeatletter
\@ifpackageloaded{caption}{}{\usepackage{caption}}
\@ifpackageloaded{subcaption}{}{\usepackage{subcaption}}
\makeatother

\usepackage{bookmark}

\IfFileExists{xurl.sty}{\usepackage{xurl}}{} 
\hypersetup{
  pdftitle={A Latent Oscillator Measurement Model to Simulate Emotional-Expression Score Dynamics in Video},
  pdfauthor={Aleksandar Tomašević; Hudson Golino; Alexander P. Christensen},
  pdfkeywords={classifier scores, simulation, dimensional recovery,
compositional data, dynamic networks},
  colorlinks=true,
  linkcolor={blue},
  filecolor={Maroon},
  citecolor={Blue},
  urlcolor={Blue},
  pdfcreator={LaTeX via pandoc}}

\title{A Latent Oscillator Measurement Model to Simulate
Emotional-Expression Score Dynamics in Video}
\author{Aleksandar Tomašević \and Hudson Golino \and Alexander P.
Christensen}
\date{}

\begin{document}
\maketitle
\begin{abstract}
Facial-expression classifiers convert video into multivariate time
series of scores with measurement error from classifiers, videos, and
recording conditions. Empirical score series cannot establish whether
the score channels reflect a smaller set of latent expressive processes
or whether an analysis would recover those processes. We introduce the
Latent Oscillator Measurement Model (LOMM), a data-generating model that
separates latent dynamics, time-varying activity, and a factor-analytic
observation model. The latent processes are damped, undamped, or
amplifying linear oscillators. LOMM generates bounded scores or
continuous indicators. Study 1 used four-fold cross-fitting with scores
from 100 MAFW videos to calibrate LOMM and evaluate generated series on
held-out videos. Median plausibility and coverage were 0.970 and 0.920
for LOMM, versus 0.510 and 0.370 for a calibrated static logistic-normal
generator. Study 2 tested whether Dynamic Exploratory Graph Analysis
(DynEGA), static EGA, GraphicalVAR, and GIMME recovered a known
dimensional structure from continuous indicators generated by LOMM. At
100 observations per clip, with failed or timed-out fits counted as
incorrect, correct-dimension recovery was 0.939 for DynEGA, 0.884 for
static EGA, 0.777 for GraphicalVAR, and 0.176 for GIMME. Replacing the
common fixed initialization with independent stationary starts for
stable dimensions and bounded independent starts for amplifying
dimensions reduced recovery for DynEGA, static EGA, and GraphicalVAR.
The primary initialization was optimistic relative to this alternative.
LOMM provides a controlled test of whether an analysis recovers a
specified latent structure before score patterns are interpreted
psychologically.
\end{abstract}

\section{Introduction}\label{introduction}

Studying how emotional expression unfolds over time requires
measurements that capture changes in visible behavior across successive
moments. Automated emotion detection makes this task tractable by
converting video into multivariate time series of facial-expression
scores: each frame yields a score for every emotion label in a chosen
set, allowing researchers to examine expressive variation at a temporal
resolution that manual coding cannot match. Earlier classifiers
restricted this analysis to categories represented in their training
data, but vision-language models such as CLIP can score labels supplied
as text (\citeproc{ref-foteinopoulou2023EmoCLIP}{Foteinopoulou \&
Patras, 2023}; \citeproc{ref-li2023CLIPER}{Li et al., 2023};
\citeproc{ref-radford2021LearningTransferable}{Radford et al., 2021}).
Interpreting the resulting series requires accounting for how those
scores are produced. They depend on the model, label set,
video-processing pipeline, image quality, and recording conditions, and
their accuracy varies with model adaptation, label choice, and
evaluation data (\citeproc{ref-foteinopoulou2023EmoCLIP}{Foteinopoulou
\& Patras, 2023}; \citeproc{ref-li2023CLIPER}{Li et al., 2023}). In a
small demonstration on FindingEmo images, zero-shot CLIP achieved 41.7\%
accuracy and a macro-F1 of 0.331
(\citeproc{ref-tomasevic2026TransforEmotion}{Tomašević et al., 2026}).
On the Image-Emotion dataset, fine-tuned Emotion-CLIP reached 67\%
accuracy and a macro-F1 of 0.64 in a reported 10-fold evaluation
(\citeproc{ref-bondielli2021CLIP}{Bondielli \& Passaro, 2021}).
Classifier scores therefore provide fallible measurements of visible
expression whose patterns can reflect expressive behavior, the
measurement process, or both. Their use in emotion research requires a
clear distinction between the visible expression being scored and the
subjective emotional experience that those scores alone cannot
establish.

This raises a psychometric question. Score channels can be analyzed as
separate variables, or they can be treated as fallible indicators of a
smaller number of latent expressive processes
(\citeproc{ref-mehu2015Emotion}{Mehu \& Scherer, 2015};
\citeproc{ref-nesselroade2002}{Nesselroade et al., 2002};
\citeproc{ref-russellCoreAffect2003}{Russell, 2003}). The number of
classifier labels does not establish how many distinct processes the
channels represent. Treating overlapping channels as separate dimensions
can overstate that number, while combining indicators of distinct
processes can conceal differences. Dimensionality assessment therefore
concerns both the number of latent processes and the assignment of
channels to those processes.

The observable associations among indicators can change even when the
generating dimensions and indicator assignments remain fixed. Dynamics,
activity, measurement error, and observation length affect how clearly
the dimensional structure appears in a recorded series. Indicators of
the same process may share little observable variation when that process
is weakly expressed or measurement error is large, and measurement error
can change the structure recovered by dynamic methods
(\citeproc{ref-kankaanpaa2026multipleindicator}{Kankaanpää et al.,
2026}). Changes in activity can also produce changes in the indicators
without a discontinuity in the underlying latent level. Evaluating
recovery therefore requires conditions in which these sources of
variation can be controlled while the generating structure remains
known.

We introduce the Latent Oscillator Measurement Model (LOMM) for this
purpose. LOMM combines the damped linear oscillator model of
intraindividual dynamics (\citeproc{ref-boker2015Adaptive}{Boker, 2015};
\citeproc{ref-boker2002Method}{Boker \& Nesselroade, 2002}) with an
activity layer and a factor-analytic observation model
(\citeproc{ref-nesselroade2002}{Nesselroade et al., 2002};
\citeproc{ref-nesselroade2008dfm}{Zhang et al., 2008}). It simulates
latent processes, specifies how strongly each process is expressed at
each observation, and maps the expressed processes onto observed
indicators with measurement error. These components and the observation
length can be varied while the generating dimensions and indicator
assignments remain fixed, allowing researchers to examine when a
specified structure is recovered and which conditions obscure it.
Empirical score series can guide dimensionality assessment and the
choice of simulation settings. Direct evaluation of recovery requires
simulated data for which the generating structure is known.

\begin{figure}

\centering{

\includegraphics[width=0.98\linewidth,height=\textheight,keepaspectratio]{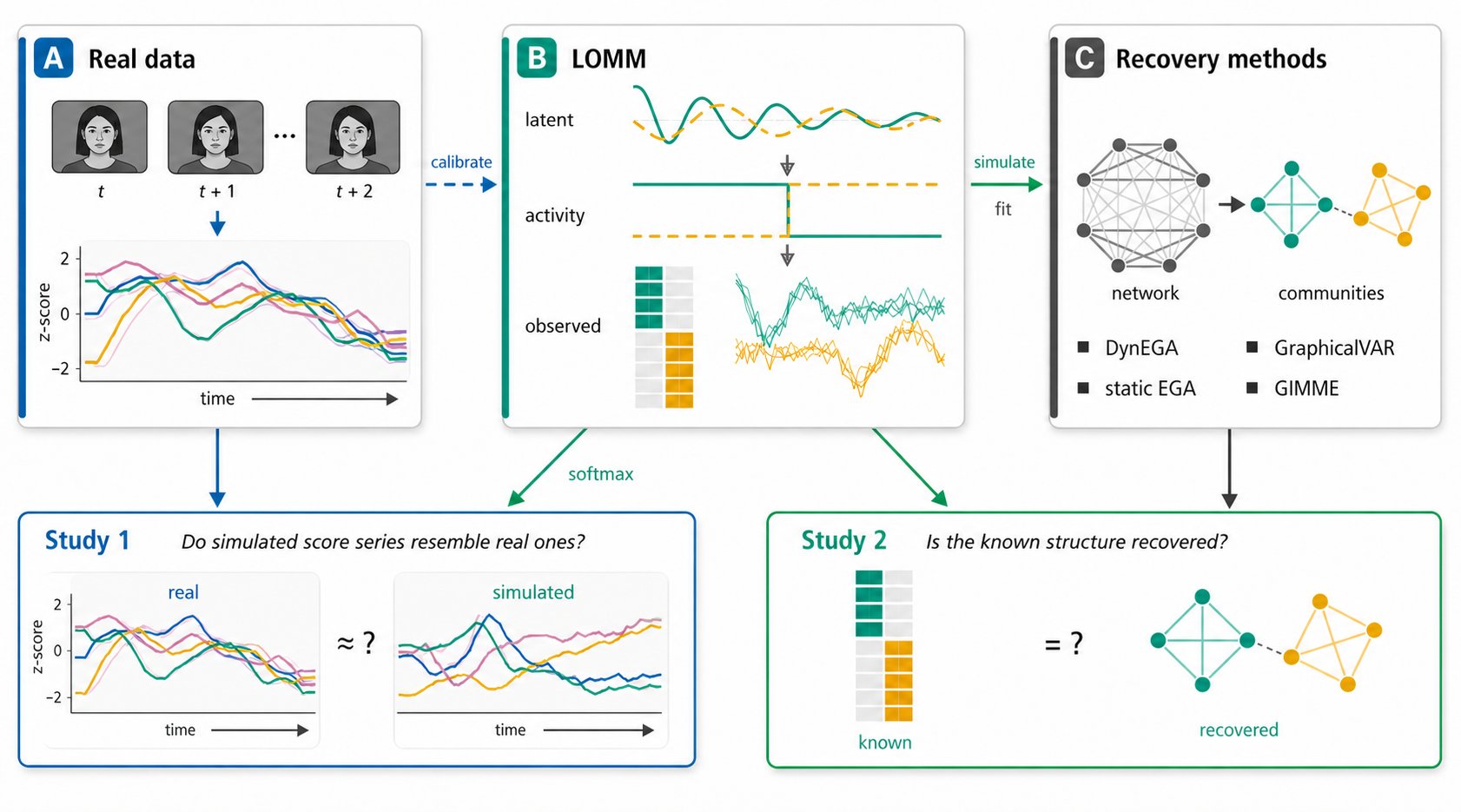}

}

\caption{\label{fig-pipeline}Overview of the two studies. (A) Real data:
a classifier scores each sampled frame of an MAFW video, giving one
score series per label. (B) LOMM: latent oscillators, an activity
schedule that sets how strongly each is expressed at each observation,
and observed indicators with a known assignment to dimensions (colored
blocks). (C) Recovery methods: DynEGA, static EGA, GraphicalVAR, and
GIMME each estimate a network among the indicators and group them into
communities. Study 1 uses MAFW calibration videos to set the generator
and compares simulated score series with MAFW evaluation series. Study 2
fits the four methods to simulated indicators and compares the recovered
communities with the known assignment.}

\end{figure}

We use LOMM in two studies (Figure~\ref{fig-pipeline}). In Study 1, we
extract emotion-score series from videos in the MAFW dataset, a
multimodal database for dynamic facial-expression recognition in the
wild (\citeproc{ref-liu_mafw_2022}{Liu et al., 2022}), and evaluate how
closely LOMM reproduces their statistical properties. Each of four folds
uses 75 MAFW videos to calibrate LOMM and 25 videos to evaluate the
generated series. A static logistic-normal generator provides a
comparison without temporal dependence. Eleven statistics computed on
scores and log-ratio coordinates define series-level plausibility and
coverage.

Study 2 evaluates recovery of a known dimensional structure from
LOMM-generated data. LOMM generates continuous indicators with a known
number of dimensions and a known assignment of indicators to dimensions,
and four network methods are used to recover this dimensional structure:
Dynamic Exploratory Graph Analysis (DynEGA), static EGA, GraphicalVAR,
and GIMME. Each method estimates a network among the indicators, which
is then partitioned into communities. The question is whether the
recovered communities match the structure that generated the data.
Static EGA ignores temporal order and is computationally inexpensive; it
shows how much structure is recoverable from contemporaneous indicator
levels alone. Study 2 also tests how dimensional recovery depends on
initialization and the transformation of continuous indicators into
bounded scores. One analysis compares the common fixed initialization
with independent stationary starts for stable dimensions and bounded
independent starts for amplifying dimensions. A second tests whether the
indicator grouping is preserved after the indicators are transformed to
bounded scores that sum to one, as classifier scores do.

\section{The Latent Oscillator Measurement
Model}\label{the-latent-oscillator-measurement-model}

Emotion labels can share broader psychological functions, including
engagement with desirable opportunities, defense against threats, and
attention to novel events (\citeproc{ref-bradley2009Orienting}{Bradley,
2009}; \citeproc{ref-lang2010Motivational}{Lang \& Bradley, 2010}).
These functions motivate an illustrative LOMM simulation with three
latent dimensions labeled approach-affiliation, threat-defense, and
orienting-novelty. Consider the threat-defense dimension: fear and
anxiety both concern responses to threat, although they involve
different defensive functions
(\citeproc{ref-mcnaughton2004Defense}{McNaughton \& Corr, 2004}). A
researcher could therefore specify fear, anxiety, worry, and
apprehension score channels as indicators of a shared threat-related
process. In the simulation, these channels covary as that process
changes over time, while each also contains measurement error. Other
channels measure the remaining two dimensions. The recovery task is to
identify the three dimensions and their indicator assignments. This
grouping is a theory-informed simulation hypothesis; whether empirical
classifier scores exhibit it requires separate investigation.

\begin{figure}

\centering{

\includegraphics[width=0.98\linewidth,height=\textheight,keepaspectratio]{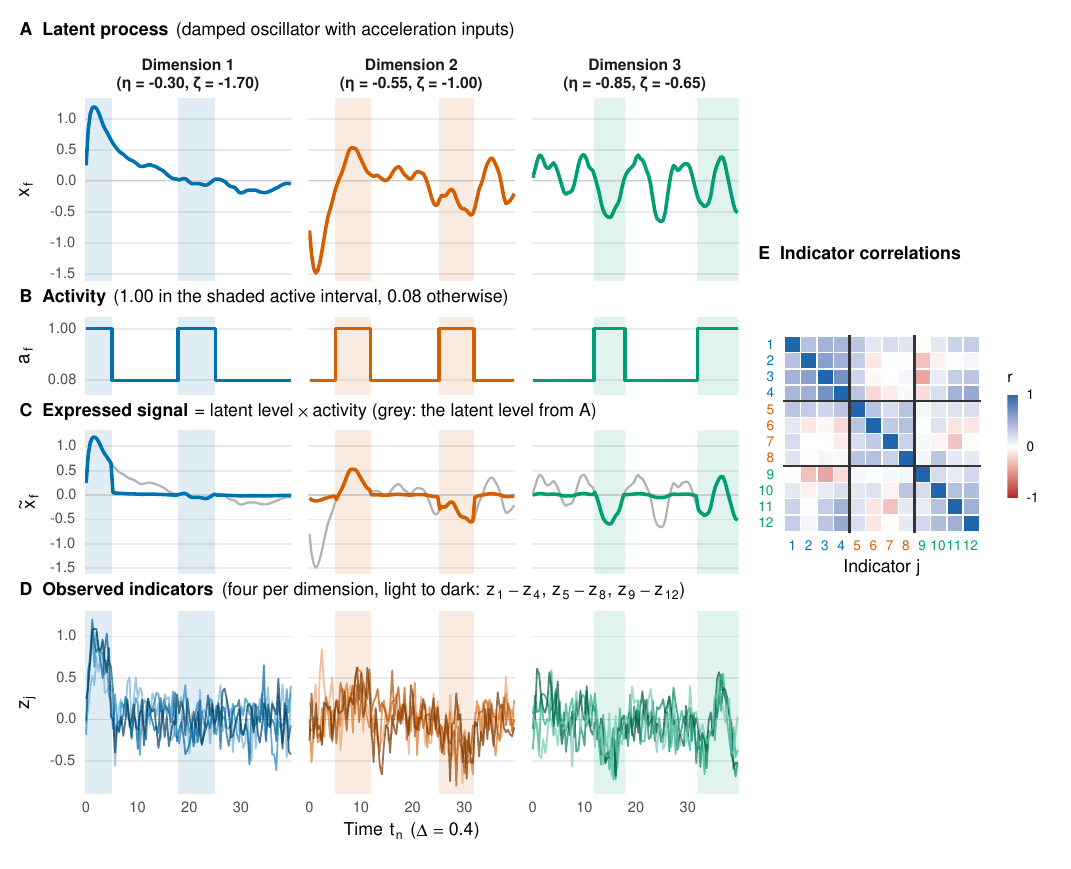}

}

\caption{\label{fig-lomm-process}One simulated LOMM clip with three
latent dimensions, \(T=100\) observations, and \(\Delta=0.4\). Each
column follows one dimension from the latent layer to the observed
indicators. (A) Latent trajectory; the oscillator parameters are given
in the column headers, and each dimension has its own initial state and
acceleration inputs. (B) Activity, 1.00 in the shaded active interval
and 0.08 otherwise. (C) Expressed signal, the product of the latent
level and the activity value, drawn over the latent level in grey. (D)
The four observed indicators of each dimension in shades from light to
dark, with primary loadings 0.65 to 0.85, secondary loadings of
\(\pm 0.30\) or \(\pm 0.35\) on another dimension for nine of the twelve
indicators, and residual SD 0.20. (E) Correlations among the twelve
indicators from this clip, ordered by dimension.}

\end{figure}

LOMM generates data in three layers (Figure~\ref{fig-lomm-process}). The
latent layer describes how each latent dimension changes over time. The
activity layer determines how strongly that dimension is expressed at
each observation. The observation layer creates multiple indicators with
prespecified loadings and measurement error. Keeping these layers
separate allows a researcher to change the dynamics, the expression
pattern, or the measurement quality while holding the other components
fixed.

\emph{Notation.} Suppose we have \(N\) video clips, each represented by
\(T\) sampled frames. Scoring each sampled frame produces one
multivariate observation. We index clips by \(i=1,\ldots,N\) and
observations within a clip by \(n=1,\ldots,T\). Observations occur at
times \(t_n=(n-1)\Delta\), where \(\Delta\) is the interval between
successive observations. Latent dimensions are indexed by
\(f=1,\ldots,m\). Continuous indicators are indexed by \(j=1,\ldots,p\);
in Study 2, \(p=mk\) with \(k\) indicators per dimension. Bounded scores
have \(J\) channels. Bold symbols are vectors or matrices.

For clip \(i\) and latent dimension \(f\), let \(x_{if}(t)\) denote the
latent level and \(\dot{x}_{if}(t)\) its rate of change. LOMM treats
each dimension as a linear second-order oscillator:

\begin{equation}
\ddot{x}_{if}(t)
=
\eta_f x_{if}(t)
+
\zeta_f\dot{x}_{if}(t)
+
q_{if,n},
\qquad t\in[t_n,t_{n+1}).
\label{eq:lomm-latent}
\end{equation}

Equation 1 is the damped linear oscillator used in research on
intraindividual dynamics (\citeproc{ref-boker2002Method}{Boker \&
Nesselroade, 2002}). Affective-dynamics models describe how emotional
states depart from and return toward a characteristic baseline. Damped
oscillator models represent this process through a tendency to return
toward baseline and a separate tendency for ongoing fluctuations to
diminish (\citeproc{ref-chow2005Thermostat}{Chow et al., 2005}). LOMM
uses these dynamics to generate latent expressive trajectories with
different patterns of persistence, recovery, and amplification.

The position-feedback parameter \(\eta_f\) determines how strongly a
deviation from baseline changes the direction of the trajectory. When
\(\eta_f<0\), values above baseline produce a downward acceleration and
values below baseline produce an upward acceleration. The
velocity-feedback parameter \(\zeta_f\) determines whether ongoing
changes are slowed or reinforced. Negative values oppose the current
direction of change and damp fluctuations; zero adds no damping;
positive values reinforce the change and amplify it. These parameters
describe the simulated dynamics; interpreting them as individual
differences in emotion regulation would require validation against
independent psychological measures. The term \(q_{if,n}\) is a random
acceleration input, drawn independently for each interval between
observations \(n\) and \(n+1\) and held constant within that interval;
it produces irregular peaks and reversals in the trajectory.

When both feedback parameters are negative, the process tends to settle
toward its baseline after a disturbance. It may approach the baseline
smoothly or move above and below it with progressively smaller swings.
New random inputs can keep these fluctuations going. Positive velocity
feedback instead reinforces ongoing changes and can produce increasingly
large swings. We include these growing trajectories to test whether the
methods can recover the specified dimensions when signal amplitudes
increase over a clip. Appendix C reports how large the simulated
amplitudes become. These conditions are evaluated over a fixed
observation period; continued growth would eventually become implausible
for sustained expressive behavior. Appendix A gives the equations used
to calculate each successive latent state, explains which parameter
settings produce stable behavior, and describes how the starting values
are chosen.

The activity layer represents how strongly a latent process contributes
to the observed expression at a particular moment. Separating a process
from its expression is motivated by evidence that outward emotional
behavior can change without a corresponding change in reported
experience, as in expressive suppression
(\citeproc{ref-gross1993Suppression}{Gross \& Levenson, 1993}). In LOMM,
the nonnegative activity value \(a_{if,n}\) scales the latent signal:
\(\tilde{x}_{if,n}=a_{if,n}x_{if,n}\). Small values weaken its
contribution to the indicators, and larger values strengthen it. The
activity schedule is specified by the researcher and can represent
periods of weaker or stronger expression. It does not identify the
psychological mechanism producing those changes. This layer matters for
recovery because an abrupt change in activity can produce a large change
in the observed indicators even when the latent level remains
continuous.

The observation layer is a standard factor model:

\begin{equation}
\mathbf z_{in}
=
\boldsymbol\Lambda_i^{(z)}\tilde{\mathbf x}_{in}
+
\boldsymbol\varepsilon_{in}^{(z)},
\qquad
\boldsymbol\varepsilon_{in}^{(z)}\sim\mathcal N(\mathbf 0,\boldsymbol\Psi_i).
\label{eq:lomm-measurement}
\end{equation}

Row \(j\) of \(\boldsymbol\Lambda_i^{(z)}\) contains the loadings of
indicator \(j\) on the \(m\) latent dimensions, and
\(\boldsymbol\varepsilon_{in}^{(z)}\) is measurement error with
covariance \(\boldsymbol\Psi_i\). Appendix A gives the complete
observation equations, including the separate notation for continuous
indicators and log-ratio coordinates and the optional clip-specific
baseline and scale terms.

Figure~\ref{fig-lomm-process} shows the three layers for one simulated
clip. In panel A the dimensions differ in their feedback parameters and
start from different states: dimension 1 rises to an early peak and
decays smoothly, dimension 2 starts in a trough and rebounds, and
dimension 3 oscillates faster and is driven mainly by its acceleration
inputs. Panel B shows the activity schedule, in which the dimensions
take turns being active. Panel C shows the product of the two layers. A
latent peak that falls inside an active interval is expressed at full
size; the same peak outside the interval is reduced to 8\% of its size
and is barely visible in the indicators. The steps at the boundaries of
the active intervals appear in the indicators although the latent level
remains continuous. Panel D shows the twelve indicators; each loads
mainly on its own dimension, and nine also carry a smaller loading on
another dimension. Panel E shows the correlation matrix among the twelve
indicators from this clip. The three diagonal blocks reflect the
indicator assignments, which is the structure that Study 2 asks the
methods to recover. The weaker correlations between blocks can reflect
both secondary loadings and sample covariation among the expressed
latent trajectories. The secondary loadings in this example are larger
than in the Study 2 cross-loading condition, which limits them to
\(\pm 0.10\).

LOMM can output two types of observed variables. The first is the vector
of continuous indicators \(\mathbf z_{in}\), used in the primary Study 2
designs; there, the factor structure that generated the data is defined
directly on the analyzed variables. The second is a vector of \(J\)
positive scores that sum to one, which matches the form of classifier
output and is used in Study 1. To produce bounded scores, LOMM maps the
latent dimensions to a \((J-1)\)-dimensional vector of log-ratio
coordinates and applies the softmax function. Because scores that sum to
one are compositional data, they are analyzed both as scores and in
log-ratio coordinates, which remove the sum constraint. A separate Study
2 condition applies the softmax directly to the \(p\) continuous
indicators and tests whether their original grouping is preserved.
Appendix A gives the softmax, log-ratio, and identification details.

In Study 1, MAFW calibration videos determine the LOMM observation-model
geometry and generator settings. The primary Study 2 design sets
between-clip baselines to zero and scales to one but lets loadings and
residual variances vary slightly across clips. Estimating oscillator,
activity, and measurement parameters from empirical data would require
the sign, scale, temperature, and frequency constraints stated in
Appendix A.

\section{Study 1: Comparing Simulated and Real Score
Series}\label{study-1-comparing-simulated-and-real-score-series}

Study 1 examined whether LOMM, after calibration to MAFW, could generate
classifier-score series with statistical properties observed in MAFW
videos excluded from calibration. The analysis used 100 videos from
MAFW, a large multimodal database of facial-expression video recorded in
naturalistic settings (\citeproc{ref-liu_mafw_2022}{Liu et al., 2022}).
Frames were scored with the transforEmotion R package
(\citeproc{ref-tomasevic2026TransforEmotion}{Tomašević et al., 2026})
using OpenAI's CLIP ViT-Large/14 model (setting \texttt{oai-large}),
with up to 50 frames sampled per video. Eleven labels from the MAFW
label set served as text prompts: anger, disgust, fear, happiness,
neutral, sadness, surprise, contempt, anxiety, helplessness, and
disappointment. For each scored frame, the eleven classifier-score
channels sum to one.

\begin{figure}

\centering{

\includegraphics[width=0.98\linewidth,height=\textheight,keepaspectratio]{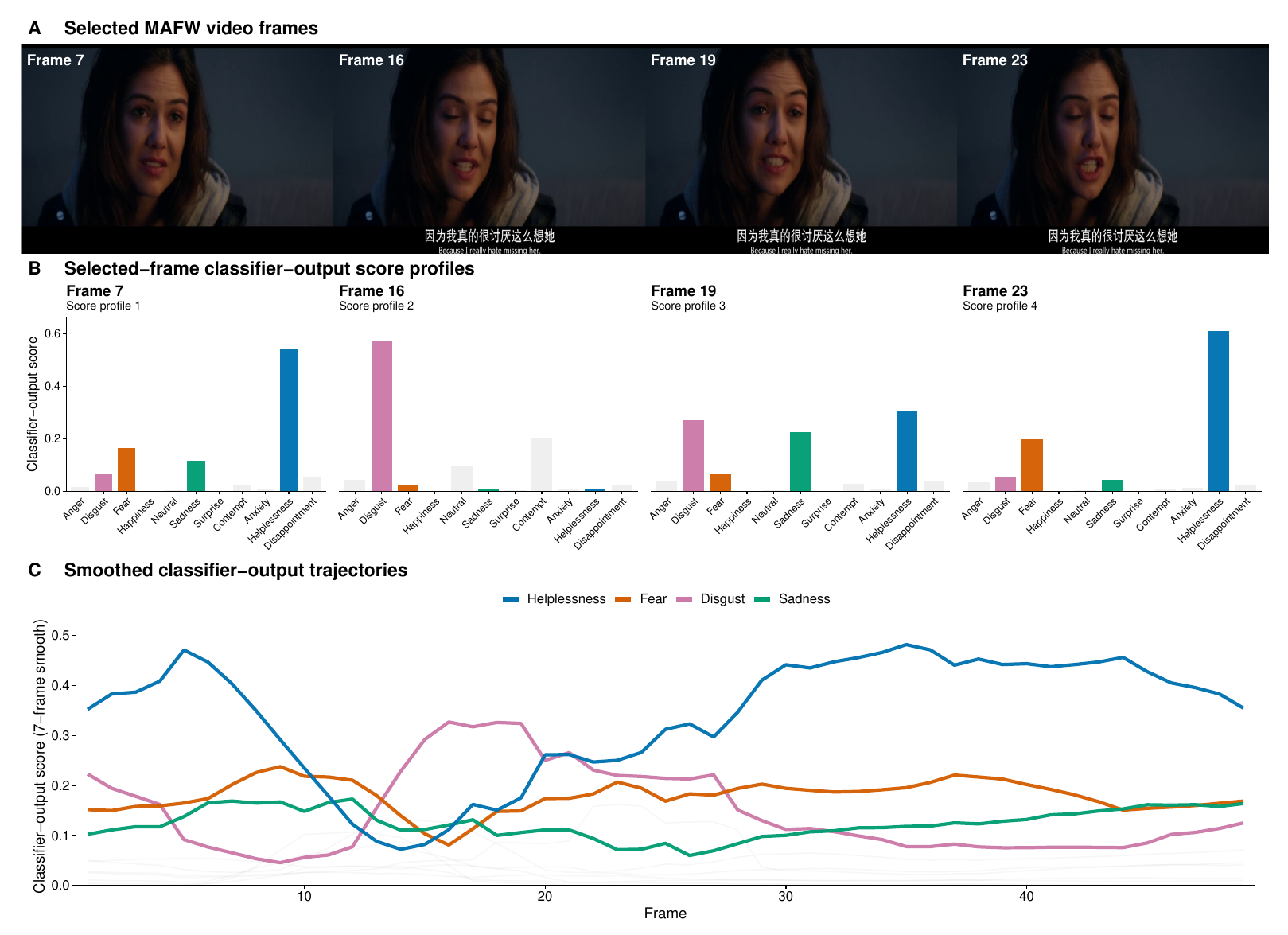}

}

\caption{\label{fig-mafwvideo}Classifier-score series from one MAFW
video. Panel A shows four frames from the video, reproduced under the
MAFW research-use terms. Panel B shows the eleven classifier scores for
those frames. Panel C shows smoothed trajectories for the channels with
the highest mean scores. Scores were generated with transforEmotion
using OpenAI CLIP ViT-Large/14.}

\end{figure}

Figure~\ref{fig-mafwvideo} shows the classifier-score series for one
MAFW video. Frames without a detected face have missing scores, so each
video was divided into segments of consecutive scored frames. The
analysis contains 3,557 observations in 188 segments. Videos contribute
16 to 50 scored observations, with a median of 35; segment lengths range
from 1 to 50 observations, with a median of 13.5. Appendix B gives the
face-detection, score-validity, missing-data, and sampling details.

We represented each MAFW and simulated series with a broad panel of
statistics covering location, dispersion, short-term change, temporal
persistence, temporal irregularity, and cross-channel dimensionality.
Eight statistics were computed within channel and video: median level;
interquartile range of levels; interquartile range of
observation-to-observation changes; lag-1 autocorrelation; and
run-length, sample, permutation, and spectral entropy. Their
channel-specific values were summarized by the median within each video.
Effective rank, robust level variance, and robust change variance
described all channels jointly. Each statistic was evaluated on the
classifier scores and a log-ratio representation, producing 22 features.
Channel statistics used centered log-ratios (CLR), and the three joint
statistics used isometric log-ratio (ILR) coordinates. The main analysis
combines the complete feature panel in the plausibility and coverage
measures. Appendix C reports the metric-specific comparisons.

We used four-fold cross-fitting. In each fold, 75 MAFW videos formed the
calibration set and 25 formed the evaluation set. The calibration set
determined the LOMM measurement geometry and the dynamic and measurement
parameters. LOMM then generated 100 datasets with the same observation
positions and uninterrupted segment boundaries as the evaluation videos.
We compared LOMM with a static logistic-normal generator calibrated from
the same 75 MAFW videos. The static generator retained the estimated
compositional structure, between-video heterogeneity, and observation
pattern while drawing observations independently over time. This
comparison estimates the gain associated with temporal generation beyond
static calibration.

For each fold, the MAFW reference comprised 1,000 comparisons between
two disjoint, category-stratified groups of 25 videos drawn from the 75
calibration videos. The 22 features were standardized using medians and
robust scales estimated from the calibration set. Series similarity was
measured by the root-mean-square Euclidean distance across these
standardized features. The similarity threshold was the 95th percentile
of the directional nearest-neighbor distances in the MAFW reference
comparisons. Plausibility is the proportion of simulated series whose
nearest MAFW evaluation series fell within this threshold. Coverage is
the proportion of MAFW evaluation series whose nearest simulation fell
within the threshold. Reported ranges are the 2.5th and 97.5th
percentiles across simulation or reference replicates. Appendix B gives
the calibration, parameter selection, metric, and reference-sampling
procedures.

\section{Study 2: Recovery of Dimensionality and Indicator
Assignments}\label{study-2-recovery-of-dimensionality-and-indicator-assignments}

\emph{Data-generating conditions.} Study 2 asks whether four
network-based methods recover the number of latent dimensions and their
indicator assignments from LOMM-generated data. The primary comparison
applies all four methods to the same simulated datasets, including a
shorter-record condition. A larger factorial simulation evaluates DynEGA
across additional combinations of dimensionality, indicators per
dimension, and observation schedules. These analyses use continuous
indicators. Each dataset contains \(N\) clips with \(T\) observations
each and \(p=mk\) indicators. All clips share the same indicator
assignments and nominal measurement parameters, while allowing small
differences in measurement strength across clips. In these comparison
and factorial designs, each primary loading receives an independent
\(\operatorname{Unif}(-0.15,0.15)\) deviation, and each residual
variance receives an independent \(\operatorname{Unif}(-0.01,0.01)\)
deviation.

A latent dimension can be present in the generating model yet contribute
little observable variation during a short recording. Weak expression
can leave its indicators dominated by measurement error. When dimensions
alternate in prominence, each contributes strongly during only part of
the recording, and transitions can change the associations among
indicators. Increasing amplitudes can make later observations contribute
disproportionately to those associations. The six scenario families
examine recovery across these different patterns of expression while
keeping the indicator assignments fixed
(Table~\ref{tbl-scenario-words}). Families A and B represent weak and
strong expression with one dimension remaining dominant. Families C and
D allow the dominant dimension to change, with more rapid switching in
D. Families E and F introduce growing amplitudes, with either one
dominant dimension or switching dominance. These families combine
activity gains, switching schedules, and dynamic settings to represent
different recovery challenges. Appendix B gives the activity schedules
and implementation details; Table B2 gives the numerical oscillator
parameters.

\begin{table}[H]

\caption{\label{tbl-scenario-words}The six Study 2 scenario families.
Numeric activity gains and dwell fractions are in Appendix B; oscillator
parameters are in Table B2.}

\centering{

\begingroup\small\setlength{\tabcolsep}{6pt}

\begin{tabular}{cl}
\toprule
Code & Family\\
\midrule
A & Weak expression, one dominant dimension\\
B & Strong expression, one dominant dimension\\
C & Strong expression, switching dominant dimension\\
D & Strong expression, rapid switching\\
E & Amplifying dynamics, one dominant dimension\\
\addlinespace
F & Amplifying dynamics, switching dominant dimension\\
\bottomrule
\end{tabular}
\endgroup

}

\end{table}%

\emph{Recovery procedures.} Network methods provide a way to estimate
both the number of dimensions and the assignment of indicators to them.
In LOMM, indicators assigned to the same latent process share a changing
signal. Under suitable measurement conditions, this shared variation can
produce groups of strongly connected indicators in an estimated network.
Community detection can therefore identify candidate dimensions without
specifying their number in advance (\citeproc{ref-golino2017ega1}{H. F.
Golino \& Epskamp, 2017}). Study 2 examines whether this correspondence
survives changes in dynamics, expression strength, measurement error,
and observation length.

The four procedures construct networks from different aspects of the
same data: indicator levels, estimated changes, or residual associations
after temporal modeling (Table~\ref{tbl-method-roles}). Every method
receives the same simulated dataset. Each estimated structure is
converted to an undirected network by a prespecified rule, and a common
community-detection step identifies groups of indicators. Comparing
these groups with the generating assignments shows how well each
representation preserves the indicator partition, while retaining each
method's statistical model and data handling.

\begin{table}[H]

\caption{\label{tbl-method-roles}The four Study 2 methods, the structure
each supplies for the common grouping step, and its purpose in the
recovery comparison.}

\centering{

\begingroup\small\setlength{\tabcolsep}{3pt}

\begin{tabular}{lp{2.35in}p{2.35in}}
\toprule
Method & Structure used for grouping & Purpose in the recovery comparison\\
\midrule
DynEGA & Partial-correlation network of estimated changes & Assess recovery from coordinated changes in the indicators\\
Static EGA & Partial-correlation network of contemporaneous levels & Assess recovery from associations among indicator levels\\
GraphicalVAR & Contemporaneous residual partial-correlation network & Assess recovery from residual associations after modeling lagged relations\\
GIMME & Average residual-covariance network across clip-level fits & Assess recovery from residual associations after fitting each clip separately\\
\bottomrule
\end{tabular}
\endgroup

}

\end{table}%

DynEGA is included to test whether indicators driven by the same latent
process can be identified through their coordinated changes over time
(\citeproc{ref-glla2010}{Boker et al., 2010};
\citeproc{ref-golino2019investigating}{H. Golino et al., 2020};
\citeproc{ref-golino2020modeling}{H. Golino et al., 2022};
\citeproc{ref-golino2017ega1}{H. F. Golino \& Epskamp, 2017}). It
estimates first derivatives from time-delay embeddings, constructs a
regularized partial-correlation network among those derivatives, and
identifies communities as candidate dimensions. GraphicalVAR
(\citeproc{ref-epskampggm}{Epskamp et al., 2018};
\citeproc{ref-graphicalVAR2024}{Epskamp, 2024}) fits a multilevel vector
autoregression that separates lagged from contemporaneous relations; its
contemporaneous residual partial-correlation network is used for
grouping.

GIMME (\citeproc{ref-gates2012group}{Gates \& Molenaar, 2012};
\citeproc{ref-lane2019uncovering}{Lane et al., 2019}) fits a structural
vector autoregression to each clip separately. This allows the fitted
temporal and contemporaneous relationships to differ across clips.
Because the generating indicator assignments are shared across clips, we
examine whether that common grouping remains detectable in the residual
associations after these separate fits. Residual covariance matrices
from successful fits are averaged and submitted to the common
community-detection procedure. This adapts GIMME's usual use for
recovering individual-specific directed relationships to the recovery of
a shared indicator partition.

Static EGA estimates a regularized partial-correlation network from the
contemporaneous correlation matrix of indicator levels, ignoring the
observation order, and applies the same community-detection step. It
serves two purposes. Statistically, it shows how much of the grouping is
recoverable from contemporaneous levels alone; if most of it is, a
dynamic model cannot add much for this recovery target. Computationally,
it requires far less time than the dynamic methods. Comparing static EGA
with DynEGA and GraphicalVAR therefore shows whether modeling temporal
structure improves grouping enough to justify its computational cost in
these conditions.

\emph{Primary design and outcomes.} The primary design uses \(T=100\)
observations per clip and contains 21,600 datasets. It crosses the six
scenario families, \(m\in\{2,3\}\) dimensions, \(N\in\{50,100\}\) clips,
three loading levels (0.40, 0.60, 0.80), three residual standard
deviations (0.125, 0.25, 0.50), and 100 replications, with \(k=4\)
indicators per dimension. Every method receives the same dataset within
a condition and replication. The empirical videos motivate an additional
short-record condition. MAFW videos contain a median of 35 scored
observations, and uninterrupted segments have a median length of 13.5
observations. We therefore also evaluate recovery with \(T=25\), using
21,600 corresponding datasets. Louvain at resolution 1.0 is the primary
community-detection algorithm; Leiden and Walktrap were applied to the
same fitted networks as sensitivity analyses.

The primary outcome is the overall correct-dimension recovery rate: the
proportion of datasets in which the method returned a valid network and
the estimated number of dimensions equaled \(m\). A failed or timed-out
fit counts as an unsuccessful recovery. We also report the valid-output
rate and the correct-dimension rate among valid outputs, so that
computational feasibility and accuracy can be examined separately.
Assignment accuracy, the adjusted Rand index, and variation of
information compare the full estimated partition of indicators with the
true partition among valid outputs. Secondary diagnostics assess
alignment between the recovered groups and the activity-scaled latent
trajectories using best-match absolute trajectory correlation and
trajectory NRMSE. Appendix B defines how these diagnostics are
constructed and matched.

\emph{Additional analyses.} Three additional analyses examine the limits
of the primary result. First, one-at-a-time sensitivity analyses start
from a reference condition (\(N=100\), \(T=100\), loading 0.60, residual
SD 0.25, common fixed initialization) and alter one design component at
a time by changing the initialization rule, smoothing activity
transitions, adding small cross-loadings, correlating residuals,
correlating acceleration inputs, or transforming the indicators to
bounded scores. The alternative initialization uses independent
stationary starts for stable dimensions and bounded independent starts
for amplifying dimensions. The bounded-score condition applies the
softmax function directly to the \(p\) continuous indicators and
analyzes either the \(p\) scores or their centered log-ratio (CLR)
components. Because the transformed scores must sum to one, the original
block structure among the \(p\) variables need not be preserved. Second,
a small analysis adds clip-specific baseline and scale differences, with
baselines constrained to the column space of the loading matrix so that
they add no structure outside the factor model. Third, an extended
factorial simulation of 518,400 datasets evaluates DynEGA alone over
additional numbers of dimensions, indicators per dimension, and
observation schedules. Appendix B gives the exact settings for every
analysis.

\section{Results}\label{results}

\subsection{Study 1 Results: Similarity of Simulated and Real Score
Series}\label{study-1-results-similarity-of-simulated-and-real-score-series}

\begin{figure}[H]

\centering{

\includegraphics[width=0.95\linewidth,height=\textheight,keepaspectratio]{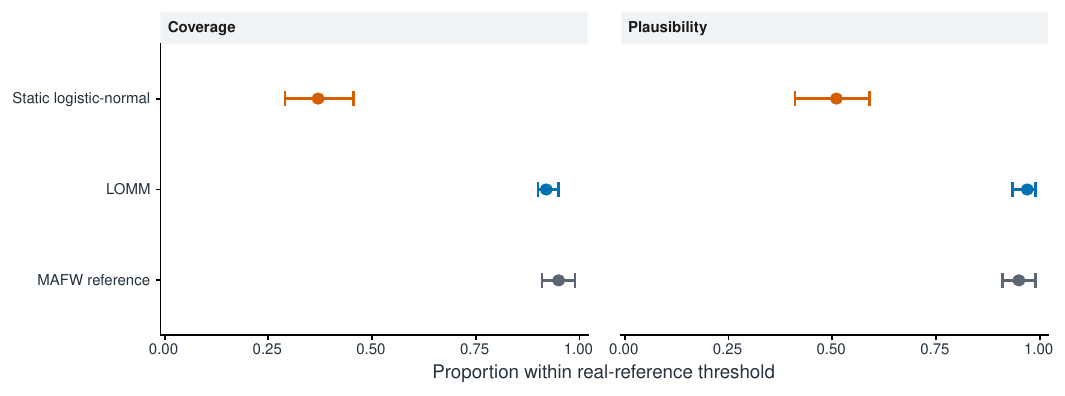}

}

\caption{\label{fig-mafw-clip-benchmark}Cross-fitted similarity of LOMM
and static logistic-normal series to MAFW evaluation series. Points are
median proportions and bars are 2.5th and 97.5th percentiles across 100
generated datasets for each generator or 1,000 MAFW-to-MAFW reference
comparisons. Plausibility is the proportion of simulated series with a
nearest MAFW evaluation series within the fold-specific threshold.
Coverage is the proportion of MAFW evaluation series with a nearest
simulation within the same threshold. The threshold is the 95th
percentile of directional nearest-neighbor distances between disjoint
MAFW calibration samples. Higher values indicate closer agreement.}

\end{figure}

Figure~\ref{fig-mafw-clip-benchmark} compares the two generators with
the MAFW reference. LOMM achieved a median plausibility of 0.970 (95\%
simulation range {[}0.935, 0.990{]}) and coverage of 0.920 {[}0.900,
0.950{]}. The corresponding MAFW-to-MAFW reference values were 0.950
{[}0.910, 0.990{]} for both measures. The static logistic-normal
generator achieved plausibility of 0.510 {[}0.410, 0.590{]} and coverage
of 0.370 {[}0.290, 0.455{]}. LOMM therefore generated score series that
usually had a close MAFW counterpart and represented most MAFW
evaluation series, whereas static calibration alone covered
substantially less of the empirical variation.

Across the metric-specific comparisons, 14 of the 22 LOMM feature
discrepancies fell within the MAFW-to-MAFW reference range. Agreement
was strongest for dispersion, short-term change, temporal persistence,
and robust change variance. The clearest remaining differences concerned
sample entropy, which was higher in LOMM series, and spectral entropy,
which was lower. Appendix C reports all metric-specific results.

\subsection{Study 2 Results: Recovery of Dimensionality and Indicator
Assignments}\label{study-2-results-recovery-of-dimensionality-and-indicator-assignments}

Table~\ref{tbl-method-overall} reports the \(T=100\) results under
Louvain. The overall recovery rate requires both a valid network and the
correct number of dimensions, so every failure and timeout counts
against it. The last two columns report assignment accuracy among valid
outputs and median runtime.

\begingroup\scriptsize\setlength{\tabcolsep}{3pt}

\begin{longtable}[]{@{}
  >{\raggedright\arraybackslash}p{(\linewidth - 10\tabcolsep) * \real{0.1287}}
  >{\raggedleft\arraybackslash}p{(\linewidth - 10\tabcolsep) * \real{0.1287}}
  >{\raggedleft\arraybackslash}p{(\linewidth - 10\tabcolsep) * \real{0.1980}}
  >{\raggedleft\arraybackslash}p{(\linewidth - 10\tabcolsep) * \real{0.1683}}
  >{\raggedleft\arraybackslash}p{(\linewidth - 10\tabcolsep) * \real{0.1980}}
  >{\raggedleft\arraybackslash}p{(\linewidth - 10\tabcolsep) * \real{0.1782}}@{}}

\caption{\label{tbl-method-overall}Results for the primary \(T=100\)
design under Louvain. Assignment accuracy is computed among valid
network outputs. Runtime is the median, in seconds, among valid network
outputs and includes model fitting, network extraction, and community
detection.}

\tabularnewline

\toprule\noalign{}
\begin{minipage}[b]{\linewidth}\raggedright
Method
\end{minipage} & \begin{minipage}[b]{\linewidth}\raggedleft
Valid output
\end{minipage} & \begin{minipage}[b]{\linewidth}\raggedleft
Correct given valid
\end{minipage} & \begin{minipage}[b]{\linewidth}\raggedleft
Overall recovery
\end{minipage} & \begin{minipage}[b]{\linewidth}\raggedleft
Assignment accuracy
\end{minipage} & \begin{minipage}[b]{\linewidth}\raggedleft
Median runtime, s
\end{minipage} \\
\midrule\noalign{}
\endhead
\bottomrule\noalign{}
\endlastfoot
DynEGA & 1.000 & 0.939 & 0.939 & 0.965 & 3.69 \\
Static EGA & 0.997 & 0.887 & 0.884 & 0.922 & 2.15 \\
GraphicalVAR & 0.941 & 0.826 & 0.777 & 0.902 & 85.41 \\
GIMME & 0.224 & 0.783 & 0.176 & 0.877 & 518.16 \\

\end{longtable}

\endgroup

DynEGA returned a valid network for all 21,600 datasets and had the
highest overall recovery rate, 0.939. Static EGA followed at 0.884,
which shows that most of the grouping was recoverable from
contemporaneous indicator levels. GraphicalVAR reached 0.777. GIMME
recovered the correct dimension in 0.783 of its valid outputs but
returned a valid network for only 0.224 of datasets, so its overall
recovery was 0.176. Assignment accuracy among valid outputs was 0.965
for DynEGA, 0.922 for static EGA, 0.902 for GraphicalVAR, and 0.877 for
GIMME. Static EGA was about 40 times faster than GraphicalVAR and had
higher overall recovery and assignment accuracy for this target.
GraphicalVAR estimates a more complex model that includes lagged
effects; the comparison evaluates only whether that additional structure
improved recovery of the grouping under these conditions. Appendix C
reports Monte Carlo standard errors, bootstrap intervals, and the
remaining partition diagnostics.

\begin{figure}[H]

\centering{

\includegraphics[width=0.98\linewidth,height=\textheight,keepaspectratio]{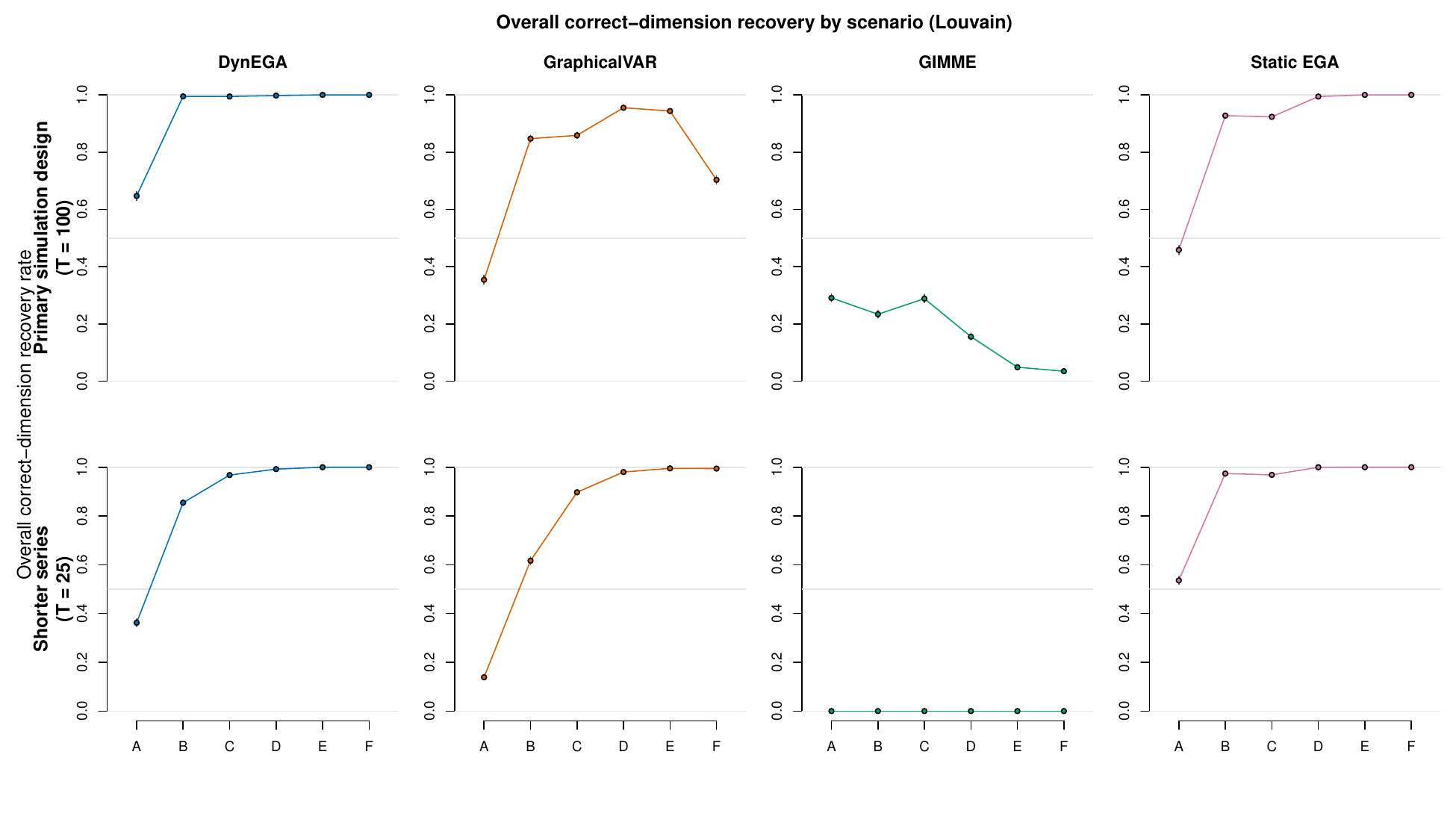}

}

\caption{\label{fig-dimension-recovery}Overall correct-dimension
recovery under Louvain by scenario family (codes as in Table 1), method,
and observations per clip (\(T=100\), top; \(T=25\), bottom). A recovery
requires a valid network output and the correct number of dimensions.
Error bars are 95\% bootstrap intervals from resampling whole datasets.}

\end{figure}%

Recovery varied across scenario families and observations per clip
(Figure~\ref{fig-dimension-recovery}). DynEGA had the highest recovery
in most conditions. Static EGA's recovery was high because the block
structure was present in contemporaneous levels. GraphicalVAR's recovery
was lowest when expression was weak or measurement error was large.
GIMME's low overall rates mainly reflected missing valid outputs. In the
extended factorial simulation, DynEGA recovered the correct number of
dimensions in 93.7\% of 518,400 datasets and assigned 95.2\% of
indicators correctly; stronger loadings and smaller residual error
improved recovery. Leiden and Walktrap led to the same conclusions as
Louvain.

Secondary diagnostics assessed alignment between the recovered groups
and the activity-scaled latent trajectories. Among valid outputs with
the correct number of dimensions at \(T=100\), the best-match absolute
trajectory correlation and trajectory NRMSE were 0.760 and 0.602 for
DynEGA, 0.791 and 0.568 for GraphicalVAR, 0.606 and 0.898 for GIMME, and
0.787 and 0.563 for static EGA. GraphicalVAR and static EGA had higher
trajectory correlations and lower NRMSE than DynEGA, while DynEGA had
higher correlation and lower NRMSE than GIMME. These descriptive
comparisons condition on the outputs satisfying each method's
dimensional recovery criterion. Grouping recovery and trajectory
recovery are different targets, and neither ordering ranks the methods
in general. Appendix C reports the complete diagnostics.

\subsection{Robustness and Sensitivity
Analyses}\label{robustness-and-sensitivity-analyses}

\emph{Shorter records.} With \(T=25\), static EGA had the highest
overall recovery, 0.913, followed by DynEGA, 0.863, and GraphicalVAR,
0.770. Shortening the record slightly reduced DynEGA's recovery, left
GraphicalVAR's about the same, and increased static EGA's. At \(T=25\),
all 21,600 GIMME attempts ended in errors and none returned a valid
network; none was recorded as a timeout. Among outputs with the correct
number of dimensions, DynEGA's trajectory correlation was 0.848 and
NRMSE was 0.521; GraphicalVAR had the highest correlation and lowest
NRMSE, followed by DynEGA and static EGA (Appendix C).

\emph{Initial states.} In the reference condition of the one-at-a-time
analyses (\(N=100\), \(T=100\), loading 0.60, residual SD 0.25, common
fixed initialization), overall recovery was 0.993 for DynEGA, 0.912 for
static EGA, and 0.683 for GraphicalVAR. Replacing the common fixed
initialization with independent stationary starts for stable dimensions
and bounded independent starts for amplifying dimensions produced the
largest change among the sensitivity conditions: recovery fell to 0.838,
0.525, and 0.253, respectively. The common fixed initialization was
therefore optimistic relative to this alternative.

\emph{Other one-at-a-time changes.} Gradual activity transitions, small
cross-loadings, and correlated acceleration inputs changed recovery
little. A residual correlation of 0.10 between every pair of indicators
increased recovery for GraphicalVAR and static EGA. This result shows
that the particular residual pattern made the target grouping easier to
recover; it does not establish robustness to residual dependence more
generally.

\emph{Bounded scores.} Transforming the \(p\) indicators to scores that
sum to one changed the variables being analyzed and sharply reduced
recovery of the original grouping. DynEGA recovered the correct number
of dimensions in 0.333 of datasets analyzed as CLR components and 0.233
analyzed as raw scores; the corresponding rates were 0.333 and 0.167 for
static EGA and 0.087 and 0.002 for GraphicalVAR. The softmax
transformation and the accompanying projection onto the zero-sum
subspace did not preserve the structure among the original indicators.
For this reason, the Study 2 recovery results apply to continuous
indicators and are not extended to bounded scores of the kind analyzed
in Study 1.

\emph{Between-clip baseline and scale.} This analysis used stable
families A through D with independent stationary initial states. Its
reference recovery rates are lower than those in analyses using the
common fixed initialization. Adding clip-specific baseline and scale
changed DynEGA's recovery from 0.775 to 0.750, a paired difference of
\(-0.025\) (95\% interval \([-0.100,0.050]\)). Static EGA rose from
0.275 to 1.000: the added baselines lie in the column space of the
loadings, so they add between-clip variation that follows the intended
grouping. GraphicalVAR and GIMME were unchanged, but those estimates
rest on very few valid outputs; GraphicalVAR produced three correct
outputs among 80 datasets, and every three-dimensional GIMME fit timed
out.

\emph{Runtime and feasibility.} Median runtime among valid \(T=100\)
outputs was 2.15 s for static EGA, 3.69 s for DynEGA, 85.41 s for
GraphicalVAR, and 518.16 s for GIMME. GraphicalVAR timed out on 1,271 of
21,600 fits. GIMME produced 4,846 valid outputs, 14,477 timeouts, and
2,277 post-estimation errors. DynEGA returned a valid output for every
fit in the primary and extended designs. These runtimes are specific to
our implementation and hardware. Appendix C gives the full runtime,
factorial, and community-detection results.

\section{Discussion}\label{discussion}

LOMM separates three determinants of multivariate time-series structure:
the dynamics of latent processes, their expression over time, and their
measurement by observed indicators. This separation matters when
researchers use associations among classifier-score channels to infer
dimensional structure. A change in the recovered grouping may reflect a
different latent structure, weaker expression of an unchanged structure,
or properties of the measurement process. LOMM allows these
possibilities to be examined under controlled conditions by holding the
generating dimensions and indicator assignments fixed while varying
dynamics, activity, or measurement quality.

The two studies establish different parts of this contribution. Study 1
supports the statistical resemblance of calibrated bounded output to
short MAFW classifier-score series. Study 2 establishes recovery
performance for specified continuous-indicator models. The
initialization and bounded-score analyses show why those recovery
results require explicit conditions: changes in the starting states or
observation transformation can substantially change the recovered
structure.

Study 1 found that LOMM generated classifier-score series with high
plausibility and broad coverage relative to the MAFW reference. The
static logistic-normal generator preserved the calibrated score geometry
and between-video heterogeneity. It omitted temporal dependence and
performed substantially worse on both measures. In this comparison, LOMM
reproduced the joint statistical profile of the MAFW series more closely
than the calibrated static logistic-normal generator. The
metric-specific results also identify a concrete target for model
development: LOMM captured most measures of level, dispersion, change,
persistence, and cross-channel variation, while its entropy profile
remained less similar to MAFW.

Study 2 evaluated recovery of a known factor structure from continuous
indicators. DynEGA had the highest overall recovery in the primary
matched-method design and was feasible at the scale of the extended
factorial simulation, where it was the only method evaluated. Static EGA
showed that most of the grouping was recoverable from contemporaneous
levels; it exceeded GraphicalVAR in both dimensional recovery and
assignment accuracy while running in a median of 2.15 s rather than
85.41 s. For this grouping target and these conditions, GraphicalVAR's
additional lagged model did not improve recovery relative to static EGA.
GraphicalVAR answers additional questions about lagged relations, which
this comparison does not evaluate.

The trajectory diagnostics in Appendix C show that the ranking of
methods depends on the recovery target. DynEGA's advantage concerned
recovery of the number of dimensions and the indicator partition;
alignment of recovered-group summaries with the latent trajectories
among valid outputs with the correct number of dimensions produced a
different ordering. The two targets should be evaluated separately. All
comparisons apply to the specific estimation, network-conversion, and
grouping procedures used here. GIMME's low overall recovery mainly
reflected its failure to return valid outputs under a clip-level fitting
procedure that differs from its usual use.

The most important qualification of the primary result concerns
initialization. Recovery was lower with independent stationary starts
for stable dimensions and bounded independent starts for amplifying
dimensions than with the common fixed initialization. The primary
initialization was therefore optimistic relative to this alternative.
Because this sensitivity changed cross-dimension and cross-clip
coherence together with the marginal initial-state distribution, it does
not isolate the effect of sharing alone. Gradual activity transitions,
small cross-loadings, and correlated acceleration inputs changed little.
The bounded-score condition set a clearer limit: the softmax
transformation changed the associations among the original indicators,
and neither the raw scores nor the CLR components preserved the original
grouping. The Study 2 recovery claims therefore apply to continuous
indicators.

\emph{Limitations.} Several limitations remain. The latent dimensions
are uncoupled. The amplifying families produce large amplitudes late in
the record and test recovery under those conditions only. The \(T=300\)
conditions also change the observation interval, so they do not isolate
the number of observations per clip. The GIMME residual-covariance
conversion was constructed for this comparison and differs from typical
idiographic use. Study 1 uses 100 short MAFW videos, one zero-shot
scoring pipeline, and one LOMM specification chosen during model
development on the same sample. LOMM is used here only as a generator
with prespecified parameters; estimating it from data would require the
identification constraints listed in Appendix A.

LOMM provides a generative basis for studying how latent dynamics,
changing expression, and measurement jointly shape multivariate
psychological time series. The present studies establish two initial
uses: generating bounded series that reproduce selected statistical
properties of video-derived emotion scores, and evaluating dimensional
recovery from continuous indicators with a known generating structure.
Together, they demonstrate how explicit assumptions about the generating
processes can support controlled evaluations of psychological
measurement. The bounded-score results also identify a central
methodological challenge: transforming latent signals into proportions
can change the observable structure on which dimensionality methods
rely.

A next step is to develop procedures for fitting LOMM to observed data
and recovering its dynamic, activity, and measurement parameters.
State-space and hierarchical continuous-time methods provide starting
points (\citeproc{ref-driver2018Hierarchical}{Driver \& Voelkle, 2018};
\citeproc{ref-ou2019Dynr}{Ou et al., 2019}), with parameter-recovery
studies needed to establish identification and estimation accuracy under
realistic sampling and measurement conditions. A particularly relevant
direction is the modeling of compositional time series, including gaze
allocation across areas of interest
(\citeproc{ref-facevicova2026Compositional}{Fačevicová et al., 2026}),
daily allocations of time among sleep and activity behaviors
(\citeproc{ref-le2022ActivitiesAffect}{Le et al., 2022}), and topic
proportions in repeated language samples
(\citeproc{ref-blei2006DynamicTopics}{Blei \& Lafferty, 2006}). In each
case, increasing one component's share necessarily reduces the combined
share of the others. LOMM's compositional observation layer offers a
basis for simulating such dependent trajectories while controlling the
latent processes that generate them. Extending this approach to
estimation would allow researchers to investigate which features of
psychological change remain identifiable when observations express
relative allocations. This would develop LOMM into a broader framework
for connecting theories of psychological dynamics with the constraints
of compositional measurement.

\newpage

\section{Declarations}\label{declarations}

\subsection{Funding}\label{funding}

The authors received no specific funding for this work.

\subsection{Competing interests}\label{competing-interests}

The authors declare that they have no competing interests.

\subsection{Ethics approval}\label{ethics-approval}

This study used simulations and secondary video data obtained under the
MAFW research-use terms. It involved no new recruitment, intervention,
or direct collection of human-participant data by the authors.

\subsection{Consent to participate}\label{consent-to-participate}

Not applicable.

\subsection{Consent for publication}\label{consent-for-publication}

The manuscript includes four still frames from MAFW solely to illustrate
the score-extraction procedure. The frames are cited to the dataset, and
no source video is redistributed. Their inclusion follows the dataset's
terms for limited reproduction in academic publications.

\subsection{Availability of data and
materials}\label{availability-of-data-and-materials}

The project GitHub repository at
\url{https://github.com/atomashevic/LOMM} contains the simulation
summary results, analysis-ready tables underlying the figures, and
aggregate MAFW-derived summaries used in the simulated-versus-real
comparison. Raw MAFW videos and frame-level classifier scores are not
redistributed. Access to MAFW is governed by the dataset provider, and
an end-to-end reanalysis of Study 1 requires approved access and local
generation of the classifier scores.

\subsection{Code availability}\label{code-availability}

Code for generating LOMM datasets, fitting the methods, reproducing the
statistical summaries, and regenerating the tables and figures is
available in the same GitHub repository. The repository also provides
the manuscript source and instructions for reproducing the analyses.

\subsection{Authors' contributions}\label{authors-contributions}

Aleksandar Tomašević contributed conceptualization, formal analysis,
methodology, resources, software, validation, visualization, writing the
original draft, and reviewing and editing the manuscript. Hudson Golino
contributed formal analysis, methodology, resources, software,
validation, visualization, writing the original draft, and reviewing and
editing the manuscript. Alexander P. Christensen contributed
methodology, software, writing the original draft, and reviewing and
editing the manuscript.

\section{Open Practices Statement}\label{open-practices-statement}

The study was not preregistered. The simulation and analysis code,
manuscript source, paper-facing tables, and aggregate MAFW-derived
summaries used in the simulated-versus-real comparison are available at
\url{https://github.com/atomashevic/LOMM}. Raw MAFW videos are
third-party materials, and the frame-level classifier scores are derived
from those restricted materials; neither is included.

\section{References}\label{references}

\begingroup
\setlength{\parindent}{-0.5in}
\setlength{\leftskip}{0.5in}

\phantomsection\label{refs}

\endgroup

\newpage

\setcounter{table}{0}
\renewcommand{\thetable}{A\arabic{table}}
\setcounter{figure}{0}
\renewcommand{\thefigure}{A\arabic{figure}}

\section{Appendix A: Complete Specification of
LOMM}\label{appendix-a-complete-specification-of-lomm}

This appendix gives the formal LOMM specification behind the conceptual
account in the main text. It is intended for readers who want to
reimplement the generator or examine its assumptions and identification
limits.

\subsection{A.1 The Latent Transition}\label{a.1-the-latent-transition}

The transition below advances the latent level and velocity over one
observation interval while holding the acceleration input constant. The
matrix \(\boldsymbol\Phi_f(\Delta)\) describes how the current state
evolves, and \(\boldsymbol\Gamma_f(\Delta)\) describes how the input
changes that state. For the oscillator in Equation 1, these are

\[
\boldsymbol{\Phi}_f(\Delta)=\exp(\mathbf{A}_f\Delta),
\qquad
\mathbf{A}_f=
\begin{bmatrix}
0 & 1 \\
\eta_f & \zeta_f
\end{bmatrix}
\qquad
\text{and}
\qquad
\boldsymbol{\Gamma}_f(\Delta)=\int_0^{\Delta}\exp(\mathbf{A}_f u)\mathbf{b}\,du,
\quad
\mathbf{b}=
\begin{bmatrix}
0 \\
1
\end{bmatrix}.
\]

For the nonzero position-feedback values used in the simulation grids,
the input map has the equivalent exact form

\[
\boldsymbol{\Gamma}_f(\Delta)
=\mathbf{A}_f^{-1}\{\boldsymbol{\Phi}_f(\Delta)-\mathbf{I}_2\}\mathbf{b}.
\]

The implemented state recurrence is

\[
\begin{bmatrix}
x_{if,n+1}\\
v_{if,n+1}
\end{bmatrix}
=
\boldsymbol{\Phi}_f(\Delta)
\begin{bmatrix}
x_{if,n}\\
v_{if,n}
\end{bmatrix}
+
\boldsymbol{\Gamma}_f(\Delta)q_{if,n},
\qquad n=1,\ldots,T-1.
\]

The integral definition remains valid when \(\mathbf{A}_f\) is singular,
including \(\eta_f=0\). If a root-based closed form is used numerically,
a separate repeated-root branch is required when \(\zeta_f^2+4\eta_f\)
is close to zero.

The state components \(x_{if,n}\) and \(v_{if,n}\) are the level and its
rate of change at \(t_n\); \(\eta_f\) is position feedback, \(\zeta_f\)
is velocity feedback, and \(\sigma_{qf}\) is the standard deviation of
the acceleration input. On the interval \([t_n,t_{n+1})\) the underlying
continuous-time model is
\(\ddot{x}_{if}(t)=\eta_f x_{if}(t)+\zeta_f\dot{x}_{if}(t)+q_{if,n}\)
with \(q_{if,n}\) held constant, so \(\boldsymbol{\Phi}_f(\Delta)\)
propagates the deterministic state and \(\boldsymbol{\Gamma}_f(\Delta)\)
loads the input.

Let \(\mathbf q_{i,n}=(q_{i1,n},\ldots,q_{im,n})^\top\) and
\(\mathbf D_q=\operatorname{diag}(\sigma_{q1},\ldots,\sigma_{qm})\). The
generator draws

\[
\mathbf q_{i,n}\sim
\mathcal N_m\!\left(\mathbf 0,\mathbf D_q\mathbf R_q\mathbf D_q\right).
\]

The primary designs use \(\mathbf R_q=\mathbf I_m\). The
correlated-input sensitivity uses
\(\mathbf R_q=(1-0.15)\mathbf I_m+0.15\mathbf 1_m\mathbf 1_m^\top\).
Table B2 reports the Study 2 values of \(\sigma_{qf}\).

In Study 1, \(\sigma_{qf}\) was set so that the stationary variance of
\(x_{if,n}\) was one. Specifically, if \(\mathbf P_f^{(1)}\) solves
\(\mathbf P_f^{(1)}=\boldsymbol\Phi_f\mathbf P_f^{(1)}\boldsymbol\Phi_f^\top+\boldsymbol\Gamma_f\boldsymbol\Gamma_f^\top\),
then
\(\sigma_{qf}=\{\mathbf e_1^\top\mathbf P_f^{(1)}\mathbf e_1\}^{-1/2}\).

Inputs are independent across clips and transitions and are independent
of initial states, activity schedules, and measurement errors. They are
also independent across dimensions when \(\mathbf R_q=\mathbf I_m\).
Each acceleration input is held constant for a full observation
interval, a convention called zero-order hold. The input is therefore
defined per interval; it is not a continuous-time white-noise process.
The covariance added in one transition is
\(\boldsymbol{\Gamma}_f(\Delta)\sigma_{qf}^{2}\boldsymbol{\Gamma}_f(\Delta)^\top\)
for a single dimension.

\begin{figure}[H]

\centering{

\includegraphics[width=0.9\linewidth,height=\textheight,keepaspectratio]{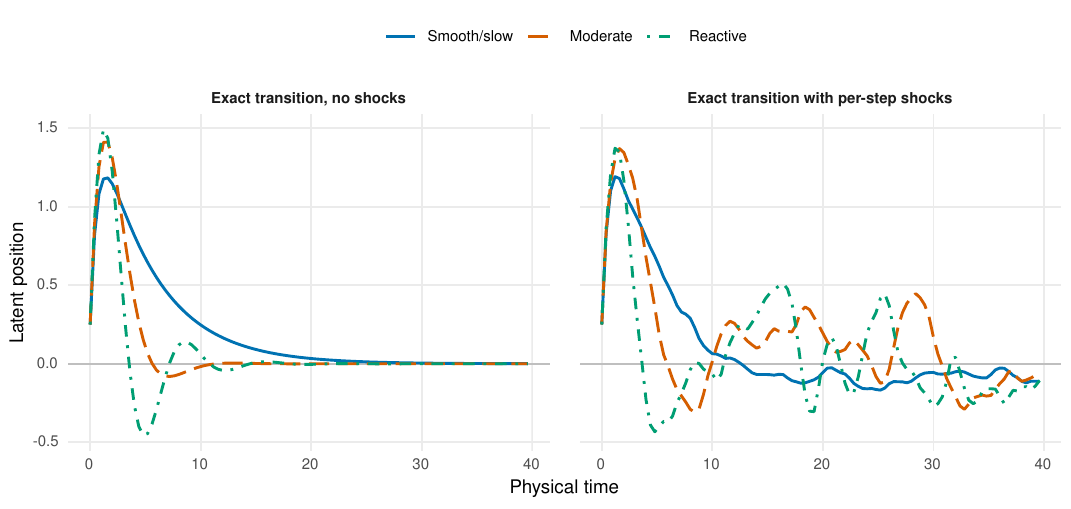}

}

\caption{\label{fig-latent-inputs}Illustrative latent trajectories for
three uncoupled dimensions under the exact transition, shown without and
with piecewise-constant acceleration inputs. The inputs create irregular
peaks and reversals while the feedback parameters govern the underlying
dynamic regime.}

\end{figure}%

\subsection{A.2 Dynamic Regimes and Finite-Horizon
Amplification}\label{a.2-dynamic-regimes-and-finite-horizon-amplification}

The characteristic roots describe whether the deterministic motion
decays, grows, or oscillates. For dimension \(f\), they are

\[
r_{f,\pm}=\frac{\zeta_f\pm\sqrt{\zeta_f^2+4\eta_f}}{2}.
\]

Asymptotic stability means that the effect of an initial state
eventually decays to zero in the absence of new inputs. In continuous
time, it requires \(\eta_f<0\) and \(\zeta_f<0\), and the roots are
oscillatory when \(\zeta_f^2+4\eta_f<0\). For the sampled transition,
the spectral radius \(\rho\) is the largest absolute eigenvalue of
\(\boldsymbol\Phi_f(\Delta)\). Stability requires

\[
\rho\{\boldsymbol{\Phi}_f(\Delta)\}
=\exp\!\left[\Delta\max\{\operatorname{Re}(r_{f,+}),\operatorname{Re}(r_{f,-})\}\right]<1.
\]

The real parts of the characteristic roots determine whether the effect
of an initial state decays or grows. Over the observation horizon
\(H=(T-1)\Delta\), we summarize this modal change by

\[
g_f(H)=\exp\!\left[H\max\{\operatorname{Re}(r_{f,+}),\operatorname{Re}(r_{f,-})\}\right].
\]

This quantity describes deterministic modal growth or decay. It does not
bound individual stochastic trajectories or capture all possible
transient amplification. Table C6 reports the maximum real part of the
roots and the modal factors, while Table C5 summarizes the amplitudes
actually produced by the generator. Activity modulation can create
additional nonstationarity and is reported separately. The amplifying
configurations are finite-record stress tests.

\subsection{A.3 Initial States}\label{a.3-initial-states}

The system is uncoupled: \(\eta_f\), \(\zeta_f\), and \(\sigma_{qf}\)
may vary by dimension, and a dimension does not directly change the
level or velocity of another dimension.

Study 1 initialized every dimension independently at the start of each
uninterrupted segment by drawing
\(\mathbf s_{if,1}=(x_{if,1},v_{if,1})^\top\) from its zero-mean
stationary Gaussian distribution. A new state was drawn after each
interruption.

Study 2's primary simulation design initialized every dimension in every
clip at \(x_{if,1}=0.25\) and \(v_{if,1}=2\). The sensitivity analysis
replaced these common fixed values with draws that were independent
across dimensions and clips.

For a stable dimension, the stationary covariance \(\mathbf P_f\) solves

\[
\mathbf{P}_f=\boldsymbol{\Phi}_f(\Delta)\mathbf{P}_f\boldsymbol{\Phi}_f(\Delta)^\top
+\boldsymbol{\Gamma}_f(\Delta)\sigma_{qf}^2\boldsymbol{\Gamma}_f(\Delta)^\top,
\]

whose solution is unique when \(\rho\{\boldsymbol{\Phi}_f(\Delta)\}<1\).
Drawing \(\mathbf{s}_{if,1}\sim\mathcal{N}(\mathbf{0},\mathbf{P}_f)\)
makes the state process stationary before the activity layer is applied;
time-varying activity modulation still renders \(a_{if}(t)x_{if}(t)\)
nonstationary.

For a dimension with positive velocity feedback, the sensitivity draws
\(x_{if,1}\sim\mathrm{Unif}(-0.25,0.25)\) and
\(v_{if,1}\sim\mathrm{Unif}(-2,2)\). The alternative therefore changes
cross-dimension and cross-clip coherence, marginal means, covariance,
stationarity, and the magnitude and direction of the initial transient;
the observed horizon is unchanged.

\subsection{A.4 The Activity Layer at a
Transition}\label{a.4-the-activity-layer-at-a-transition}

Changes in an observed indicator can arise from changes in the latent
level or changes in its activity multiplier. For the expressed signal
\(\tilde{x}_{if,n}=a_{if,n}x_{if,n}\),

\[
\frac{\tilde{x}_{if,n+1}-\tilde{x}_{if,n}}{\Delta}
=a_{if,n+1}\frac{x_{if,n+1}-x_{if,n}}{\Delta}
+\frac{a_{if,n+1}-a_{if,n}}{\Delta}x_{if,n}.
\]

The first term is latent change expressed at the new activity level. The
second is change caused by the activity multiplier. An abrupt activity
transition can therefore contribute strongly to an estimated derivative
even though the latent level remains continuous. The gradual-transition
sensitivity analysis examines whether recovery depends on these abrupt
changes. At each observation, activity multiplies the completed latent
state at the same observation index.

The activity schedule takes two nonnegative values,
\(a_{\mathrm{high}}\) and \(a_{\mathrm{low}}\). In non-switching
families the dominant dimension remains at \(a_{\mathrm{high}}\) for the
whole record and the others remain at \(a_{\mathrm{low}}\). In switching
families, dwell fractions determine how long each dimension remains
dominant before the schedule reassigns \(a_{\mathrm{high}}\). Appendix B
lists the numeric gains and dwell values used in the Study 2 families.

\subsection{A.5 The Observation Layer}\label{a.5-the-observation-layer}

LOMM has two observation models. The continuous-indicator model maps
latent processes directly to observed variables with specified loadings
and measurement error. The bounded-score model first generates log-ratio
coordinates and then converts them to positive scores that sum to one.
Its ILR coordinates combine the logarithms of multiple score channels;
there is no separate ILR coordinate for each channel. We use separate
notation for continuous indicators, \(\mathbf z\), and log-ratio
coordinates, \(\mathbf u\).

Let \(J\) denote the number of positive score channels, let \(d=J-1\),
and let \(\mathbf V\in\mathbb R^{J\times d}\) be a fixed orthonormal
isometric log-ratio (ILR) basis satisfying
\(\mathbf V^\top\mathbf V=\mathbf I_d\) and
\(\mathbf V^\top\mathbf 1_J=\mathbf 0\). The Study 1 basis used here is
a normalized Helmert basis constructed in the stated 11-channel order.
The full observation model in log-ratio coordinates is

\[
\mathbf u_{in}
=
\boldsymbol\mu^{(u)}+\mathbf b_i^{(u)}
+\kappa s_i^{(u)}
\left(
\boldsymbol\Lambda_i^{(u)}\tilde{\mathbf x}_{in}
+\boldsymbol\varepsilon_{in}^{(u)}
\right).
\]

Here
\(\mathbf u_{in},\boldsymbol\mu^{(u)},\mathbf b_i^{(u)},\boldsymbol\varepsilon_{in}^{(u)}\in\mathbb R^d\),
\(s_i^{(u)}>0\) is the clip scale, \(\kappa>0\) is a global scale, and
\(\boldsymbol\Lambda_i^{(u)}\in\mathbb R^{d\times m}\) maps the \(m\)
activity-scaled latent positions into the log-ratio coordinates. The
residual covariance is
\(\boldsymbol\Sigma_i^{(u)}\in\mathbb R^{d\times d}\), with
\(\boldsymbol\varepsilon_{in}^{(u)}\mid\boldsymbol\Sigma_i^{(u)}\sim\mathcal N(\mathbf 0,\boldsymbol\Sigma_i^{(u)})\).
The clip scale multiplies both signal and residual, and the baseline is
additive.

The continuous-indicator branch has output dimension \(p\) and uses
separate measurement parameters:

\[
\mathbf z_{in}
=
\mathbf b_i^{(z)}
+s_i^{(z)}
\left(
\boldsymbol\Lambda_i^{(z)}\tilde{\mathbf x}_{in}
+\boldsymbol\varepsilon_{in}^{(z)}
\right),
\qquad
\boldsymbol\varepsilon_{in}^{(z)}\mid\boldsymbol\Psi_i
\sim\mathcal N(\mathbf 0,\boldsymbol\Psi_i),
\]

where
\(\mathbf z_{in},\mathbf b_i^{(z)},\boldsymbol\varepsilon_{in}^{(z)}\in\mathbb R^p\),
\(\boldsymbol\Lambda_i^{(z)}\in\mathbb R^{p\times m}\),
\(\boldsymbol\Psi_i\in\mathbb R^{p\times p}\), and \(s_i^{(z)}>0\). In
both branches, residuals are conditionally independent over time and
across clips and are independent of latent states, initial states,
activity schedules, and acceleration inputs. Designs with fixed
measurement parameters use a common loading matrix and residual
covariance within the relevant branch.

The residual-correlation sensitivity condition applies to the continuous
indicators and sets

\[
\boldsymbol\Psi
=s_\varepsilon^2
\left\{(1-\rho)\mathbf I_p
+\rho\mathbf 1_p\mathbf 1_p^\top\right\},
\qquad \rho=0.10,
\]

which introduces residual covariance between every pair of the \(p\)
indicators, including pairs from different prespecified blocks.

The baseline constructions are also branch-specific. Study 1 draws
\(\mathbf b_i^{(u)}\) from a zero-mean Gaussian whose covariance is
estimated from MAFW calibration videos in the full \(d\)-dimensional
log-ratio coordinate space. Study 2's between-clip-variability analysis
uses

\[
\mathbf b_i^{(z)}
=\boldsymbol\Lambda^{(z)}\boldsymbol\beta_i,
\qquad E(\boldsymbol\beta_i)=\mathbf 0,
\]

so that between-clip variation stays within the column space of the
continuous-indicator loading matrix. Study 2 conditions without baseline
and scale differences set \(\mathbf b_i^{(z)}=\mathbf 0\) and
\(s_i^{(z)}=1\), giving

\[
\mathbf z_{in}
=\boldsymbol\Lambda_i^{(z)}\tilde{\mathbf x}_{in}
+\boldsymbol\varepsilon_{in}^{(z)}.
\]

This equation defines the continuous indicators used in Study 2's
primary recovery designs before any bounded-score mapping. Each of the
\(p=mk\) indicators loads on one dimension and has diagonal measurement
error. In the matched-method and extended factorial designs, each
primary loading receives an independent
\(\operatorname{Unif}(-0.15,0.15)\) deviation from its nominal value,
and each residual variance receives an independent
\(\operatorname{Unif}(-0.01,0.01)\) deviation from \(s_\varepsilon^2\).
The one-at-a-time sensitivity and between-clip-variability designs use
fixed equal values except where the stated condition changes them. Zero
baseline and unit scale remove baseline and overall-scale differences;
common loadings and residual covariance require the separate
fixed-parameter condition.

For bounded output, LOMM converts the coordinates to strictly positive
scores that sum to one, the open probability simplex. The centered
log-ratio (CLR) expresses each channel relative to the geometric mean of
all channels. The isometric log-ratio (ILR) represents the same
information in \(d=J-1\) orthonormal coordinates. Formally,

\[
\operatorname{clr}(\mathbf p)
=\log\mathbf p-
J^{-1}\{\mathbf 1_J^\top\log(\mathbf p)\}\mathbf 1_J,
\qquad
\operatorname{ilr}_{\mathbf V}(\mathbf p)
=\mathbf V^\top\operatorname{clr}(\mathbf p),
\]

with the logarithm applied componentwise. For the canonical bounded LOMM
output, applying softmax to \(\mathbf V\mathbf u_{in}/\tau\) gives

\[
\operatorname{ilr}_{\mathbf V}(\mathbf p_{in})=\frac{\mathbf u_{in}}{\tau},
\qquad
\operatorname{clr}(\mathbf p_{in})
=\frac{\mathbf V\mathbf u_{in}}{\tau},
\]

The second identity follows because the components of
\(\mathbf V\mathbf u_{in}\) sum to zero. The fixed basis excludes the
unidentified common-logit direction and the softmax produces strictly
positive rows summing to one, so the scores identify the scaled
coordinates \(\mathbf u_{in}/\tau\) exactly. A common-logit component
remains unobservable, and the unscaled coordinates and \(\tau\) are not
separately identified without an external scale constraint.

The Study 2 bounded-score sensitivity condition uses the \(p=mk\)
continuous-indicator vector directly. With \(J=p\), it computes

\[
\mathbf p^{\mathrm{dep}}_{in}
=\operatorname{softmax}\left(\frac{\mathbf z_{in}}{\tau}\right),
\qquad
\operatorname{clr}\!\left(\mathbf p^{\mathrm{dep}}_{in}\right)
=\frac{1}{\tau}\mathbf P_p\mathbf z_{in},
\qquad
\mathbf P_p
=\mathbf I_p-\frac{\mathbf 1_p\mathbf 1_p^\top}{p}.
\]

For any orthonormal ILR basis
\(\mathbf V_p\in\mathbb R^{p\times(p-1)}\),
\(\mathbf V_p\mathbf V_p^\top=\mathbf P_p\). Defining
\(\mathbf u^{\mathrm{dep}}_{in}=\mathbf V_p^\top\mathbf z_{in}\)
therefore gives
\(\mathbf V_p\mathbf u^{\mathrm{dep}}_{in}=\mathbf P_p\mathbf z_{in}\).
Softmax is invariant to the removed row mean, so the direct-softmax
branch is equivalent to applying the basis form to
\(\mathbf u^{\mathrm{dep}}_{in}\). The analysis used the \(p\) raw
scores and the \(p\) CLR components, which satisfy a zero-sum
constraint; it did not analyze the \((p-1)\) ILR coordinates. The
common-logit direction in \(\mathbf z_{in}\) is unidentifiable after
this transformation.

In the equal-loading reference condition used for this sensitivity, an
equal increase in all expressed latent levels adds the same amount to
every logit. Softmax removes this common component. Specifically,
\(\boldsymbol\Lambda^{(z)}\mathbf 1_m=\lambda\mathbf 1_p\), so
\(\mathbf P_p\boldsymbol\Lambda^{(z)}\mathbf 1_m=\mathbf 0\). The
projected loading matrix \(\mathbf P_p\boldsymbol\Lambda^{(z)}\) has
rank \(m-1\), leaving one fewer independent latent direction. The
transformation therefore changes the latent representation as well as
the associations among indicators. The original channel grouping may
still contain recoverable information; this sensitivity analysis tests
whether methods recover that grouping after the transformation.

\subsection{A.6 Converting Period and Damping to Feedback
Parameters}\label{a.6-converting-period-and-damping-to-feedback-parameters}

The Study 1 calibration search specified dynamics by natural period and
damping ratio. The following equations convert those settings to the
feedback parameters \(\eta_f\) and \(\zeta_f\). If \(P_f\) is the
natural period in observation intervals and \(\xi_f\) is the damping
ratio, then

\[
\omega_{n,f}=\frac{2\pi}{P_f\Delta},
\qquad
\eta_f=-\omega_{n,f}^2,
\qquad
\zeta_f=-2\xi_f\omega_{n,f},
\]

and the damped angular frequency is
\(\omega_{d,f}=\omega_{n,f}\sqrt{1-\xi_f^2}\). With the Study 1 settings
\(P_f=12\), \(\xi_f=0.20\), and \(\Delta=1\), these give
\(\eta_f=-0.2742\) and \(\zeta_f=-0.2094\). The acceleration inputs were
scaled to give unit stationary position variance.

\subsection{A.7 Identifiability}\label{a.7-identifiability}

The simulations fix the generating parameters in advance. Estimating
them from observed data would require additional constraints because
different parameter combinations can produce the same observations. The
main ambiguities concern scale, dimension labels, and sampling
frequency.

For either observation branch \(h\in\{u,z\}\) and any nonsingular
diagonal matrix \(\mathbf{D}\),

\[
\boldsymbol{\Lambda}^{(h)}\tilde{\mathbf{x}}_{in}
=(\boldsymbol{\Lambda}^{(h)}\mathbf{D}^{-1})(\mathbf{D}\tilde{\mathbf{x}}_{in}),
\]

so loading magnitude and latent-trajectory scale are not jointly
identified without an explicit scale constraint. Activity gain and
acceleration-input scale contribute to latent amplitude in the same way.
The softmax layer adds invariance to a common location shift and a
tradeoff between the temperature and the coordinate scale. Dimension
labels are recoverable only up to permutation.

Sampling creates another ambiguity: different oscillation frequencies
can produce the same values at the observation times. This is called
aliasing. For an underdamped dimension, write

\[
r_{f,\pm}=\frac{\zeta_f}{2}\pm i\omega_f,
\qquad
\omega_f=
\frac{1}{2}\sqrt{-(\zeta_f^2+4\eta_f)}.
\]

At observation times \(t_n=(n-1)\Delta\), the frequencies \(\omega_f\)
and \(\omega_f+2\pi q/\Delta\) for integer \(q\) generate the same
sampled sine and cosine sequences after an appropriate change in initial
velocity. Identifying continuous-time oscillator parameters from data
therefore requires a cadence-dependent restriction such as

\[
0<\omega_f<\frac{\pi}{\Delta},
\]

together with a specified acceleration-input model and constraints on
the latent state basis. Together, these ambiguities require sign and
scale constraints on the dimensions, normalization of the activity
schedule, anchoring of either the loading scale or the input scale, a
reference-logit or sum-to-zero constraint, a fixed or otherwise
identified temperature, and this frequency restriction.

Every oscillatory mode used here satisfies the frequency restriction. In
Study 1, the selected damped angular frequency is approximately 0.513
radians per observation; the largest frequency in the calibration search
is approximately 0.524. Both are below the sampling bound of \(\pi\).
Study 2 uses \(\Delta\in\{0.25,0.4\}\), giving bounds of approximately
12.57 and 7.85 against a largest angular frequency of approximately
1.04. The simulations do not estimate any of these parameters.

\newpage

\setcounter{table}{0}
\renewcommand{\thetable}{B\arabic{table}}
\setcounter{figure}{0}
\renewcommand{\thefigure}{B\arabic{figure}}

\section{Appendix B: Study Designs and Implementation
Details}\label{appendix-b-study-designs-and-implementation-details}

This appendix gives the design and implementation details for both
studies. Study 1 evaluates resemblance to real bounded-score series.
Study 2 evaluates recovery of a prespecified factor structure from
continuous indicators.

\subsection{B.1 Study 1 Scoring and
Calibration}\label{b.1-study-1-scoring-and-calibration}

The sample contains 100 MAFW videos selected across the available
expression categories. Category labels served only to stratify sampling
and fold construction. The analysis retained 3,557 complete observations
in 188 uninterrupted segments. Individual videos contain 16 to 50
complete observations, with a median of 35. Segment lengths range from 1
to 50 observations, with a median of 13.5.

transforEmotion sampled up to 50 frames uniformly from each video and
used OpenAI CLIP ViT-Large/14. A Haar cascade detected faces, and the
largest detected face was expanded by 50 pixels on each side before
classification. Frames without a detected face received missing scores.
Eleven lowercase labels were passed directly to the text encoder: anger,
disgust, fear, happiness, neutral, sadness, surprise, contempt, anxiety,
helplessness, and disappointment. Burned-in subtitles were not masked
when they fell inside the padded face crop.

Scores were required to lie in \([0,1]\). A complete row was retained
when its sum was positive, finite, and within \(10^{-6}\) of one, then
divided by that sum in the fixed channel order. Values were not
truncated, and zero replacement was disabled. The transformation stops
with an error if a complete row contains a zero. None of the 39,127
retained scores equaled zero; the minimum before row normalization was
approximately \(2.42\times10^{-5}\). The analysis used complete
observations and preserved the uninterrupted segment boundaries.

\emph{Features and segment boundaries.} The evaluation panel contains
eleven statistics: median level, interquartile range of levels,
interquartile range of changes, lag-1 autocorrelation, run-length
entropy, sample entropy, permutation entropy, spectral entropy,
effective rank, robust level variance, and robust change variance. The
eight univariate statistics were computed for each of the eleven score
channels and each of their eleven CLR coordinates. The three joint
statistics used the score panel and its ten ILR coordinates. For each
video, the median of the finite channel values summarized each
univariate statistic. Together with the joint statistics, these
summaries give 22 features. A feature remained undefined if no channel
had a finite value.

Levels were pooled across all retained observations within a video.
Changes and lagged pairs were formed within segments and then pooled
across the video; no pair crossed a segment boundary. Lag-1 correlation
was computed from these pooled pairs. Run lengths were formed separately
within segments, after removing zero changes, capped at eight, and
pooled across segments. Sample and permutation entropy pooled windows
formed entirely within segments. Sample entropy compared these valid
windows across the video, including windows from different segments.
Spectral entropy was computed separately for each eligible segment and
averaged using segment length minus one as the weight. A segment
containing one observation contributed to level statistics but supplied
no changes, lagged pairs, or entropy windows.

Change IQR and robust change variance required at least one
within-segment change. Lag-1 correlation required at least two pairs and
nonzero variation in the first and second observations in the pairs.
Run-length entropy required four observations across the video and was
zero when fewer than two runs remained.

Sample entropy used order 2, windows of three observations, and a
tolerance of 0.20 times the channel's standard deviation across the
video. It required at least two windows and used the negative logarithm
of \((A+0.5)/(B+1)\), where \(A\) counts matching pairs of
three-observation windows and \(B\) counts matches between their first
two observations; a zero or undefined channel standard deviation
produced zero. Permutation entropy used order 3, required at least two
windows, and resolved ties by position within the window.

Spectral entropy required segments of at least eight observations with
positive total spectral power and at least two positive frequency
powers. Statistics failing their requirements remained undefined.

Effective rank used the positive eigenvalues of the video correlation
matrix and required that matrix to be finite. Robust level variance was
the 10\% trimmed mean squared distance from the video center divided by
the coordinate dimension. Robust change variance was the 10\% trimmed
mean squared within-segment first difference divided by twice the
coordinate dimension. The joint statistics required at least two
observations.

\emph{Calibration.} The 100 videos were divided into four
category-stratified folds. Each fold used 75 calibration videos and 25
evaluation videos. The calibration videos determined the coordinate
center, loading subspace, between-video baseline covariance, and
empirical distribution of within-video scale. The first three
eigenvectors of the pooled within-video covariance supplied the loading
directions. The between-video baseline covariance was shrunk halfway
toward its diagonal, and its eigenvalues were floored at \(10^{-8}\).
The generator included between-video baseline and scale variation.
Activity modulation was disabled: every latent dimension had an activity
multiplier of one at every generated observation.

Generator settings were selected within each calibration set. The
dynamic search crossed periods of 12, 24, and 48 observations with
damping ratios of 0.01, 0.05, and 0.20. Selection used the interquartile
range of changes, lag-1 autocorrelation, run-length entropy, and
spectral entropy. Conditional on the selected dynamics, the measurement
search crossed residual standard deviations of 0.10, 0.20, and 0.30 with
\(\kappa\) values of 0.75, 1.00, and 1.25. Measurement selection used 16
discrepancy components: median level, level IQR, change IQR, lag-1
autocorrelation, run-length entropy, spectral entropy, effective rank,
and total variance, each for scores and log-ratios. Total variance was
the sum of the coordinate variances. The additional entropy and robust
variance statistics belonged to the subsequent 22-feature evaluation.
Each candidate was evaluated with three calibration simulations. All
four folds selected a period of 12 observations, damping ratio 0.20,
residual standard deviation 0.30, and \(\kappa=0.75\).

The static logistic-normal generator used the same calibrated coordinate
center, covariance geometry, between-video baseline and scale
distributions, observation positions, and segment boundaries. It drew
independent log-ratio observations within each video. Each selected
generator produced 100 datasets per fold with 25 simulated series in
each dataset.

\emph{Evaluation.} The MAFW reference used 1,000 comparisons per fold.
Each comparison drew two disjoint, category-stratified groups of 25
videos from the 75 calibration videos. Every feature was standardized by
the median and scale of its finite values in the calibration videos. The
scale was the IQR, with successive fallbacks to 1.4826 times the median
absolute deviation, the standard deviation, and one; a data-derived
scale had to exceed \(10^{-8}\). Before standardization, any undefined
video feature was replaced by its calibration median and therefore
became zero after standardization. This replacement used no evaluation
videos. One MAFW video had no segment of eight observations, leaving its
two spectral-entropy features undefined; these were the only two
undefined values among the 2,200 empirical video features. The distance
between two series was the root-mean-square Euclidean distance across
all 22 standardized features, including features replaced in this way.
The fold-specific similarity threshold was the 95th percentile of the
pooled directional nearest-neighbor distances between the two MAFW
groups.

For each generated dataset, plausibility was the proportion of its 25
simulated series whose nearest MAFW evaluation series fell within the
threshold. Coverage was the proportion of the 25 MAFW evaluation series
whose nearest simulated series fell within the threshold. Results were
averaged across the four folds. The reported ranges are the 2.5th and
97.5th percentiles across the 100 generated datasets or 1,000 MAFW
reference comparisons.

\emph{Distributional comparisons.} Calibration and metric-specific
distributional discrepancy used finite channel-by-video contributions
for each univariate statistic, and one contribution per video for each
joint statistic. For univariate statistics, these distributions pooled
channel values across videos. Undefined contributions were omitted
separately from the empirical and simulated distributions. Their
absolute median difference and one-dimensional Wasserstein distance were
divided by the calibration scale, using the same scale fallback rule,
and the larger standardized value was retained. The measurement
objective averaged its 16 components, giving equal weight to scores and
log-ratios; an entirely undefined component made the full selection loss
infinite. All stored calibration objectives were finite. The reported
full distributional discrepancy averaged the 22 evaluation components.

\subsection{B.2 Study 2 Simulation
Designs}\label{b.2-study-2-simulation-designs}

For each clip, LOMM sets or draws the starting state and acceleration
inputs according to the configuration, advances each latent dimension,
applies the activity schedule, and constructs the observed indicators.
The primary, shorter-series, and extended factorial designs use
continuous indicators. All clips share the same indicator assignments
and nominal measurement parameters. The generator draws clip-specific
loading and residual-variance perturbations as specified in the
random-draw table below. The one-at-a-time sensitivity conditions use
fixed equal measurement parameters except for the stated change.

\begin{table}[H]

\caption{\label{tbl-scenario-families}The six Study 2 scenario families.
Dwell values are fractions of the number of observations per clip.}

\centering{

\begingroup\scriptsize\setlength{\tabcolsep}{3pt}

\begin{tabular}{cp{1.75in}cccp{1.25in}}
\toprule
Code & Family & Switching & Active / background gain & Dwell fraction & Dynamic rule\\
\midrule
A & Weak expression, one dominant dimension & No & 0.35 / 0.05 & 1.00 / 1.00 & Stable, low input scale\\
B & Strong expression, one dominant dimension & No & 1.00 / 0.08 & 1.00 / 1.00 & Stable, standard input scale\\
C & Strong expression, switching dominant dimension & Yes & 1.00 / 0.08 & 0.20 / 0.45 & Stable, standard input scale\\
D & Strong expression, rapid switching & Yes & 1.25 / 0.08 & 0.03 / 0.10 & Stable, high input scale\\
E & Amplifying dynamics, one dominant dimension & No & 1.00 / 0.08 & 1.00 / 1.00 & Positive velocity feedback\\
\addlinespace
F & Amplifying dynamics, switching dominant dimension & Yes & 1.10 / 0.10 & 0.12 / 0.30 & Positive velocity feedback\\
\bottomrule
\end{tabular}
\endgroup

}

\end{table}%

In non-switching scenarios, one dimension remains dominant for a whole
clip and the dominant dimension rotates across clips. Switching
scenarios sample dwell intervals and change the dominant dimension over
time. Families A through D use stable dynamics. Families E and F use
positive velocity feedback and test recovery under finite-horizon
amplification.

The dynamic parameters are listed in Table B2. Each vector gives one
value per latent dimension. The primary and shorter designs use
\(\Delta=0.4\); the extended design with 300 observations uses
\(\Delta=0.25\). The displayed \(\sigma_q\) values are input standard
deviations per transition. Holding them fixed while changing \(\Delta\)
changes the input specification per unit of time.

\begingroup\scriptsize\setlength{\tabcolsep}{4pt}
\begin{table}[H]

\caption{Dynamic parameters for each scenario family. Entries in each vector follow the dimension order. Acceleration-input SD is the per-step standard deviation $\sigma_q$. Four-dimensional conditions occur only in the extended factorial simulation.}
\centering
\footnotesize
\setstretch{1}

\begin{tabular}[t]{ccp{1.70in}p{1.70in}p{1.65in}}
\toprule
Code & $m$ & $\eta$ & $\zeta$ & Acceleration-input SD\\
\midrule
A & 2D & -0.30; -0.55 & -1.70; -1.00 & 0.06; 0.15\\
A & 3D & -0.30; -0.55; -0.85 & -1.70; -1.00; -0.65 & 0.06; 0.15; 0.23\\
A & 4D & -0.30; -0.55; -0.85; -1.10 & -1.70; -1.00; -0.65; -0.30 & 0.06; 0.15; 0.23; 0.33\\
B & 2D & -0.30; -0.55 & -1.70; -1.00 & 0.12; 0.30\\
B & 3D & -0.30; -0.55; -0.85 & -1.70; -1.00; -0.65 & 0.12; 0.30; 0.45\\
\addlinespace
B & 4D & -0.30; -0.55; -0.85; -1.10 & -1.70; -1.00; -0.65; -0.30 & 0.12; 0.30; 0.45; 0.65\\
C & 2D & -0.30; -0.55 & -1.70; -1.00 & 0.12; 0.30\\
C & 3D & -0.30; -0.55; -0.85 & -1.70; -1.00; -0.65 & 0.12; 0.30; 0.45\\
C & 4D & -0.30; -0.55; -0.85; -1.10 & -1.70; -1.00; -0.65; -0.30 & 0.12; 0.30; 0.45; 0.65\\
D & 2D & -0.30; -0.55 & -1.70; -1.00 & 0.15; 0.38\\
\addlinespace
D & 3D & -0.30; -0.55; -0.85 & -1.70; -1.00; -0.65 & 0.15; 0.38; 0.56\\
D & 4D & -0.30; -0.55; -0.85; -1.10 & -1.70; -1.00; -0.65; -0.30 & 0.15; 0.38; 0.56; 0.81\\
E & 2D & -0.05; -0.10 & 0.08; 0.10 & 0.25; 0.25\\
E & 3D & -0.05; -0.10; -0.15 & 0.08; 0.10; 0.12 & 0.25; 0.25; 0.25\\
E & 4D & -0.05; -0.10; -0.15; -0.20 & 0.08; 0.10; 0.12; 0.14 & 0.25; 0.25; 0.25; 0.25\\
\addlinespace
F & 2D & -0.05; -0.10 & 0.08; 0.10 & 0.35; 0.35\\
F & 3D & -0.05; -0.10; -0.15 & 0.08; 0.10; 0.12 & 0.35; 0.35; 0.35\\
F & 4D & -0.05; -0.10; -0.15; -0.20 & 0.08; 0.10; 0.12; 0.14 & 0.35; 0.35; 0.35; 0.35\\
\bottomrule
\end{tabular}
\end{table}\endgroup

The primary \(T=100\) design contains 21,600 datasets: 216 factorial
conditions with 100 paired replications per condition. The factorial
crosses six scenario families, \(m\in\{2,3\}\) dimensions,
\(N\in\{50,100\}\) clips, loadings \(\{0.40,0.60,0.80\}\), residual
standard deviations \(\{0.125,0.25,0.50\}\), and four indicators per
dimension. The shorter-series design generates 21,600 corresponding
datasets with \(T=25\), \(\Delta=0.4\), and duration 9.6. Every method
receives the same dataset within each condition and replication.

The one-at-a-time sensitivity analyses fix \(N=100\), \(T=100\), loading
0.60, residual standard deviation 0.25, and four indicators per
dimension (the reference condition). They cross all six families, two
dimensionalities, and 50 paired replications while changing one feature
at a time.

\begin{table}[H]

\caption{\label{tbl-challenge-variants}One-at-a-time sensitivity
conditions. Each changes one design component of the reference
condition.}

\centering{

\begingroup\scriptsize\setlength{\tabcolsep}{3pt}

\begin{tabular}{p{1.55in}p{4.75in}}
\toprule
Sensitivity condition & Specification\\
\midrule
Alternative initialization & Stationary zero-mean starts for stable dimensions; bounded independent starts for amplifying dimensions\\
Gradual activity transitions & A five-observation transition replaces each abrupt switch\\
Cross-loadings & Normal cross-loadings with pre-truncation SD 0.05, truncated to [-0.10, 0.10]\\
Residual equicorrelation & Correlation 0.10 between every residual pair, including pairs from different dimensions\\
Correlated inputs & Acceleration-input correlation 0.15 between dimensions\\
\addlinespace
Bounded scores & Direct softmax of the p indicators with tau = 1, analyzed as raw scores or CLR components\\
\bottomrule
\end{tabular}
\endgroup

}

\end{table}%

\begin{table}[H]

\caption{\label{tbl-random-draws}Random draws and perturbations.}

\centering{

\begingroup\scriptsize\setlength{\tabcolsep}{3pt}

\begin{tabular}{p{1.10in}p{1.25in}p{3.95in}}
\toprule
Component & Scope & Specification\\
\midrule
Acceleration input & Studies 1 and 2 & $\mathbf q_{i,n}\sim\mathcal N_m(\mathbf 0,\mathbf D_q\mathbf R_q\mathbf D_q)$. The primary designs use $\mathbf R_q=\mathbf I_m$; the correlated-input sensitivity sets every off-diagonal correlation to 0.15. Draws are independent across clips and transitions. Study 1 scales $\sigma_{qf}$ to give unit stationary position variance; Table B2 gives the Study 2 values.\\
Initial state & Study 1 & At the start of each uninterrupted segment, each dimension is drawn independently as $\mathbf s_{if,1}\sim\mathcal N_2(\mathbf 0,\mathbf P_f)$, where $\mathbf P_f$ is the stationary covariance for the selected transition and input scale.\\
Activity & Study 1 & Activity modulation is disabled: $a_{if,n}=1$ for every dimension and generated observation.\\
Primary loadings & Study 2 matched and extended designs & For item $j$ assigned to dimension $f$, $\Lambda_{i,jf}^{(z)}=\lambda+U_{ij}$ with independent $U_{ij}\sim\operatorname{Unif}(-0.15,0.15)$. All cross-loadings are zero.\\
Residual variances & Study 2 matched and extended designs & For indicator $j$, $\Psi_{i,jj}=s_\varepsilon^2+V_{ij}$ with independent $V_{ij}\sim\operatorname{Unif}(-0.01,0.01)$; off-diagonal entries are zero.\\
\addlinespace
Switching dwell & Study 2 switching families & Set $L_{\min}=\max\{1,\lfloor Td_{\min}\rfloor\}$ and $L_{\max}=\max\{L_{\min},\lceil Td_{\max}\rceil\}$. Each dwell is sampled uniformly from the integers $L_{\min},\ldots,L_{\max}$; the final dwell is truncated at observation $T$. The first dominant dimension rotates across clips, and each later dominant dimension is sampled uniformly from the other $m-1$ dimensions.\\
Cross-loadings & Study 2 sensitivity & Each non-primary loading is drawn independently from $\mathcal N(0,0.05^2)$ and truncated componentwise to $[-0.10,0.10]$. Primary loadings remain fixed at 0.60.\\
Factor baseline $\boldsymbol\beta_i$ & Study 2 between-clip analysis & $\boldsymbol\beta_i\sim\mathcal N_m(\mathbf 0,1.8487025519^2\mathbf I_m)$ and $\mathbf b_i^{(z)}=\boldsymbol\Lambda^{(z)}\boldsymbol\beta_i$.\\
Clip scale $s_i^{(z)}$ & Study 2 between-clip analysis & $s_i^{(z)}\sim\operatorname{Unif}(0.5,1.5)$, independently across clips. It multiplies the within-clip signal and measurement residual before the baseline is added.\\
\bottomrule
\end{tabular}
\endgroup

}

\end{table}%

The between-clip analysis uses 80 paired datasets from stable families A
through D, \(m\in\{2,3\}\), and 10 replications. Each dataset has
\(N=100\), \(T=100\), four indicators per dimension, loading 0.60,
residual standard deviation 0.25, and independent stationary starts. Its
baseline varies within the loading space according to
\(\boldsymbol\beta_i\sim\mathcal N_m(\mathbf 0,1.8487025519^2\mathbf I_m)\),
and its clip scale follows
\(s_i^{(z)}\sim\operatorname{Unif}(0.5,1.5)\).

The extended factorial simulation evaluates DynEGA on 518,400 datasets.
It crosses \(N\in\{50,100\}\), \(T\in\{100,300\}\), \(m\in\{2,3,4\}\),
\(k\in\{4,6\}\), three loading levels, three residual-error levels, six
scenarios, and 400 replications. The \(T=100\) conditions use
\(\Delta=0.4\) and the \(T=300\) conditions use \(\Delta=0.25\) while
holding per-step input variance fixed. Their difference therefore
combines the number of observations, the observation interval, the
physical duration, and the discrete input specification.

The public lomm\_observe() function implements the observation step and
records the configuration, seeds, and hashes needed for reproduction.
Study 2 uses the benchmark generator documented in the analysis code,
which returns continuous indicators for the primary designs and applies
the direct-softmax and CLR transformation for the bounded-score
sensitivity condition.

\subsection{B.3 Method
Implementations}\label{b.3-method-implementations}

DynEGA estimates first derivatives from time-delay embeddings,
constructs an EBICglasso partial-correlation network, and groups its
positive edges (\citeproc{ref-glla2010}{Boker et al., 2010};
\citeproc{ref-golino2019investigating}{H. Golino et al., 2020};
\citeproc{ref-golino2020modeling}{H. Golino et al., 2022};
\citeproc{ref-golino2017ega1}{H. F. Golino \& Epskamp, 2017}). Video
identity prevents embedding windows from crossing clip boundaries. An
explicit search selects the embedding dimension by the Total Entropy Fit
Index, with shorter embeddings breaking ties. Candidate dimensions are
\(\{5,10,15,25,50\}\) for \(T=100\) and \(\{5,6,10,12\}\) for \(T=25\).
EGAnet 2.3.0 uses GLLA embedding parameters tau = 1 and delta = 1, first
derivatives, automatic correlations, pairwise missing-data handling, and
the glasso model. The derivative scale does not affect the correlation
matrix because conversion from observation units to physical time
multiplies every derivative by the same positive constant.

Static EGA applies EBICglasso to the contemporaneous correlation matrix
of the stacked indicator levels and then uses the same
network-processing and community-detection rules as DynEGA. It ignores
the observation order. This makes it a benchmark that shows how much of
the grouping is recoverable without temporal information, and a
computational benchmark for the temporal procedures.

GraphicalVAR uses mlGraphicalVAR with clip identity as the grouping
variable, within-clip centering, variable standardization, and gamma 0.5
(\citeproc{ref-epskampggm}{Epskamp et al., 2018};
\citeproc{ref-graphicalVAR2024}{Epskamp, 2024}). Lagged pairs never
cross clip boundaries. The grouping uses the returned fixedPCC
contemporaneous residual partial-correlation matrix. The diagonal is
removed and the matrix is treated as an undirected weighted graph.

GIMME fits each standardized clip separately with unified structural
equation modeling and vector autoregression
(\citeproc{ref-gates2012group}{Gates \& Molenaar, 2012};
\citeproc{ref-lane2019uncovering}{Lane et al., 2019}). Standardized
residual covariance matrices from successful fits are aligned by
variable name, averaged across clips, and symmetrized. The resulting
graph is a residual-covariance graph. Unequal residual variances can
therefore influence its weights. This conversion was built for the
present comparison and differs from typical idiographic use.

For every method, negative graph weights are set to zero. If no positive
edge remains, absolute values of the original off-diagonal weights are
used; only an all-zero matrix produces singleton communities. The
integrated outputs do not record the frequency of this fallback. Louvain
at resolution 1.0 is primary (\citeproc{ref-blondel2008fast}{Blondel et
al., 2008}). Leiden and Walktrap are applied to the same fitted network
as sensitivity analyses (\citeproc{ref-pons2006walktrap}{Pons \& Latapy,
2005}; \citeproc{ref-traag2019leiden}{Traag et al., 2019}). A
deterministic offset from the dataset seed initializes each method fit.

GraphicalVAR and GIMME were omitted from the 518,400-dataset extended
factorial simulation because the full design was computationally
impractical. Both were included in the primary and shorter-series
designs. GraphicalVAR was also included in the one-at-a-time sensitivity
and between-clip analyses. GIMME was included in the between-clip
analysis and excluded from the one-at-a-time sensitivity analyses.

\subsection{B.4 Outcomes, Uncertainty, and
Computing}\label{b.4-outcomes-uncertainty-and-computing}

A method can fail to recover the generating dimensions because it
returns no usable network or because its network yields the wrong number
of groups. The primary recovery rate counts both outcomes as failures
and uses all attempted datasets as its denominator. We also report the
proportion of attempts that return a valid network and the proportion of
valid outputs with the correct number of dimensions. These supporting
rates separate computational failure from incorrect grouping.

Adjusted Rand index and variation of information compare the full
estimated and prespecified partitions among valid outputs. Higher
adjusted Rand index and lower variation of information indicate closer
agreement. Assignment accuracy matches estimated groups to dimensions to
maximize correct indicator assignments. Every indicator remains in the
denominator, and indicators in unmatched groups count as incorrect.

Trajectory diagnostics assess whether the recovered groups summarize the
time course of the generating processes. They are constructed from the
levels of the analyzed indicators after grouping, including for DynEGA.
Scores retain the original observation order within each clip. The
indicators are standardized over the pooled clips, and each recovered
group is summarized by its first principal component. A single-indicator
group uses its standardized indicator, and a row mean is used if a
component cannot be computed. Each resulting summary is then
standardized over the pooled clips. Let \(\mathbf y_{ig}\) denote the
summary of recovered group \(g\) in clip \(i\), and let
\(\tilde{\mathbf x}_{if}\) denote the activity-scaled trajectory of
data-generating dimension \(f\). The best-match absolute trajectory
correlation is

\[
R_{\mathrm{best}}
=
\frac{1}{Nm}
\sum_{i=1}^{N}\sum_{f=1}^{m}
\max_g
\left|
\operatorname{cor}
\left(\tilde{\mathbf x}_{if},\mathbf y_{ig}\right)
\right|.
\]

Correlations use pairwise complete observations. The maximizing
recovered group is selected separately for each data-generating
dimension, so the same recovered-group summary may be selected for
multiple dimensions.

Normalized root mean squared error (NRMSE) measures the difference
between the standardized trajectories. It uses a separate one-to-one
assignment within each clip. For data-generating dimension \(f\) and
recovered group \(g\), the sign-invariant cost is

\[
d_{ifg}
=
\min_{\delta\in\{-1,1\}}
\left[
\frac{1}{K_{ifg}}
\sum_{n\in\mathcal O_{ifg}}
\left\{
\frac{\tilde{x}_{if,n}-\bar{x}_{if}}{s_{x,if}}
-\delta
\frac{y_{ig,n}-\bar{y}_{ig}}{s_{y,ig}}
\right\}^{2}
\right]^{1/2},
\]

where \(\mathcal O_{ifg}\) contains the \(K_{ifg}\) observations for
which both standardized series are finite. The means and sample standard
deviations are computed separately for each series within each clip. Let
\(\mathcal O_{x,if}\) contain the \(K_{x,if}\) finite observations of
the activity-scaled latent trajectory. Its sample variance is

\[
s_{x,if}^{2}
=
\frac{1}{K_{x,if}-1}
\sum_{n\in\mathcal O_{x,if}}
\left(\tilde{x}_{if,n}-\bar{x}_{if}\right)^2,
\]

and \(s_{y,ig}^{2}\) is defined analogously for the recovered-group
summary. Thus, NRMSE is the root mean squared difference between two
within-clip standardized series after choosing the sign that gives the
smaller value.

For each clip, a linear assignment algorithm selects the one-to-one
matching \(\widehat\pi_i\) that minimizes the total cost. A comparison
with fewer than three finite pairs after standardization, including
comparisons involving a constant series, receives a cost of 2. The cost
matrix is padded with a value of 2 when a data-generating dimension has
no recovered-group match. The dataset-level diagnostic is

\[
\operatorname{NRMSE}
=
\frac{1}{Nm}
\sum_{i=1}^{N}\sum_{f=1}^{m}
d_{if,\widehat\pi_i(f)}.
\]

Correlation allows several latent dimensions to share the same recovered
summary; NRMSE assigns a different summary to each matched dimension.
All-attempt failure-adjusted versions assign correlation 0 and NRMSE 2
when no valid network output exists or the corresponding dataset
diagnostic is undefined.

Uncertainty uses 2,000 bootstrap resamples of whole datasets within each
reported analysis group. Monte Carlo standard errors use
delete-one-dataset jackknifing. The three community-detection algorithms
reuse the same fitted structure and therefore assess the grouping step
rather than estimator variability.

Analyses used R 4.5.1 on Rocky Linux 8.8 with dual Intel Xeon Silver
4310 processors, 48 logical threads, and 125 GiB RAM. Key versions were
EGAnet 2.3.0, graphicalVAR 0.3.4, gimme 10.0, igraph 2.2.1, qgraph
1.9.8, psych 2.5.6, lavaan 0.6-20, dplyr 1.1.4, and ggplot2 4.0.1.
GraphicalVAR and GIMME used a 900-second limit per fit in the primary
and shorter designs.

Runtime is elapsed wall-clock time for attempts returning a valid
network output. It includes structure estimation, network extraction,
and community detection. One fitted structure is reused for Louvain,
Leiden, and Walktrap, so feasibility counts unique fits and accuracy
counts fit-and-algorithm results. Runtime values are specific to this
implementation and machine.

\section{Appendix C: Supplementary
Results}\label{appendix-c-supplementary-results}

\setcounter{table}{0}
\renewcommand{\thetable}{C\arabic{table}}
\setcounter{figure}{0}
\renewcommand{\thefigure}{C\arabic{figure}}

\subsection{C.1 Metric-Level Results for Study
1}\label{c.1-metric-level-results-for-study-1}

\begin{figure}[H]

\centering{

\includegraphics[width=0.95\linewidth,height=\textheight,keepaspectratio]{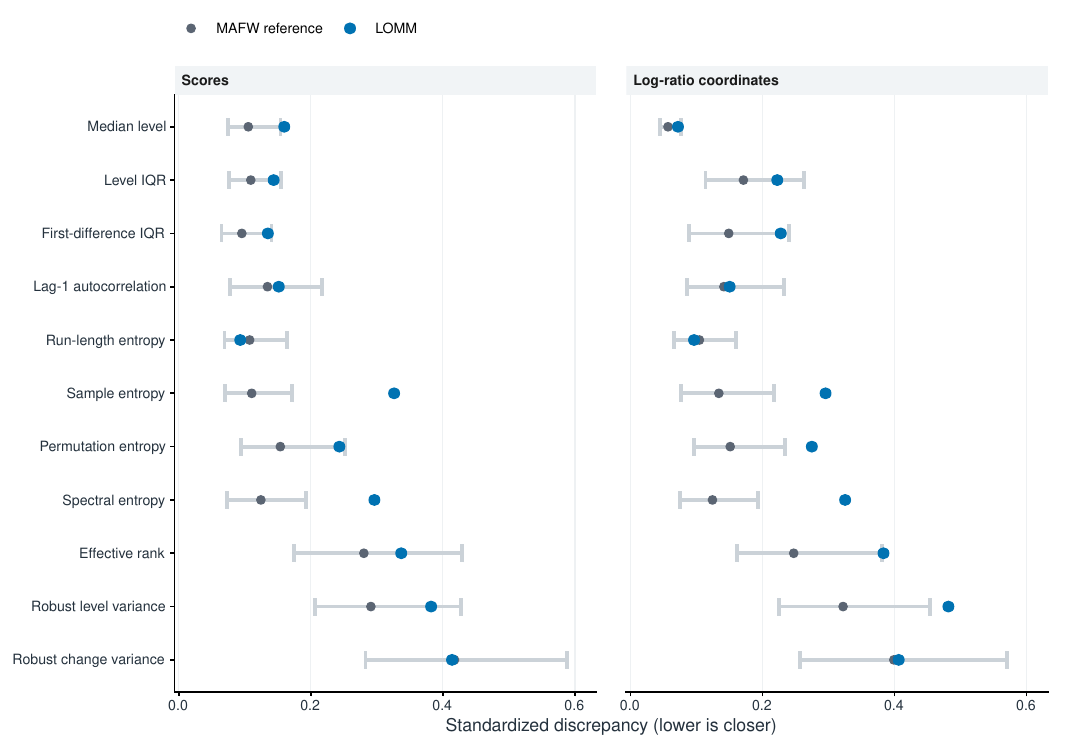}

}

\caption{\label{fig-mafw-metric-benchmark}Metric-level discrepancies
between LOMM and MAFW classifier-score series. Blue points are medians
across 100 LOMM simulations. Gray points and bars show the median and
central 95\% reference range across 1,000 disjoint MAFW-to-MAFW
comparisons. Results are averaged across four folds and shown for scores
and log-ratio representations. Discrepancy is the larger of the
standardized absolute median difference and standardized Wasserstein
distance. Smaller values indicate closer agreement.}

\end{figure}

Figure~\ref{fig-mafw-metric-benchmark} compares LOMM with the
MAFW-to-MAFW reference for each of the 22 features. Fourteen LOMM
discrepancies fell within the reference range. The full 22-feature
distributional discrepancy was 0.2543 {[}0.2356, 0.2738{]} for LOMM and
0.1811 {[}0.1521, 0.2193{]} for the MAFW reference. The corresponding
median for the static logistic-normal generator was 0.6942.

LOMM closely reproduced most measures of level, dispersion, short-term
change, persistence, effective dimensionality, and robust variation. Its
largest departures occurred for entropy. Sample entropy was higher and
spectral entropy was lower than in MAFW for both scores and log-ratio
coordinates. Log-ratio permutation entropy was lower, effective rank was
higher, and robust level variance was lower. The score median-level
discrepancy was just outside the reference range, although the signed
median difference was 0.004.

Diagnostic subsets gave the same interpretation. A time-domain subset
containing level, spread, change, lag-1 autocorrelation, and run-length
entropy had a LOMM discrepancy of 0.1463, within the MAFW reference
interval {[}0.0967, 0.1489{]}. A local-temporal subset containing
change, lag-1 autocorrelation, and run-length entropy also fell within
its reference interval. The two robust variance measures together fell
within the MAFW reference range. The entropy measures accounted for most
of the remaining full-panel discrepancy.

\newpage

\subsection{C.2 Grouping and Trajectory
Diagnostics}\label{c.2-grouping-and-trajectory-diagnostics}

This appendix reports two types of post-estimation diagnostic. Grouping
diagnostics evaluate recovered indicator groupings among valid network
outputs. Trajectory diagnostics assess alignment between recovered-group
summaries and activity-scaled latent trajectories; the trajectories are
not direct latent-state estimates produced by the methods.

\begin{landscape}\scriptsize\setlength{\tabcolsep}{1.25pt}
\begin{table}

\caption{Louvain grouping diagnostics for the matched-method simulation with $T=100$. All diagnostics use valid network outputs. Signed dimension error is the estimated number of dimensions minus the generating number; absolute error is its magnitude. Higher adjusted Rand index and assignment accuracy, and lower variation of information, indicate closer agreement. MCSE denotes Monte Carlo standard error. The 95\% intervals resample whole datasets within each reported analysis group.}
\centering
\begin{tabular}[t]{llrrcrrr}
\toprule
Method & Diagnostic & Estimate & MCSE & 95\% interval & Datasets & Valid outputs & Correct dimensions\\
\midrule
DynEGA & Signed dimension error & 0.166 & 0.0057 & {}[0.154, 0.177] & 21600 & 21600 & 20282\\
DynEGA & Absolute dimension error & 0.168 & 0.0057 & {}[0.157, 0.180] & 21600 & 21600 & 20282\\
DynEGA & Adjusted Rand index & 0.935 & 0.0015 & {}[0.932, 0.938] & 21600 & 21600 & 20282\\
DynEGA & Variation of information & 0.095 & 0.0023 & {}[0.091, 0.100] & 21600 & 21600 & 20282\\
DynEGA & Assignment accuracy & 0.965 & 0.0009 & {}[0.963, 0.966] & 21600 & 21600 & 20282\\
\addlinespace
GraphicalVAR & Signed dimension error & 0.979 & 0.0171 & {}[0.947, 1.011] & 20329 & 20329 & 16784\\
GraphicalVAR & Absolute dimension error & 0.979 & 0.0171 & {}[0.945, 1.012] & 20329 & 20329 & 16784\\
GraphicalVAR & Adjusted Rand index & 0.867 & 0.0022 & {}[0.863, 0.872] & 20329 & 20329 & 16784\\
GraphicalVAR & Variation of information & 0.183 & 0.0031 & {}[0.177, 0.189] & 20329 & 20329 & 16784\\
GraphicalVAR & Assignment accuracy & 0.902 & 0.0016 & {}[0.899, 0.906] & 20329 & 20329 & 16784\\
\addlinespace
GIMME & Signed dimension error & 0.319 & 0.0110 & {}[0.298, 0.341] & 4846 & 4846 & 3795\\
GIMME & Absolute dimension error & 0.319 & 0.0110 & {}[0.298, 0.341] & 4846 & 4846 & 3795\\
GIMME & Adjusted Rand index & 0.733 & 0.0057 & {}[0.722, 0.744] & 4846 & 4846 & 3795\\
GIMME & Variation of information & 0.337 & 0.0073 & {}[0.323, 0.351] & 4846 & 4846 & 3795\\
GIMME & Assignment accuracy & 0.877 & 0.0027 & {}[0.872, 0.882] & 4846 & 4846 & 3795\\
\addlinespace
Static EGA & Signed dimension error & 0.799 & 0.0159 & {}[0.770, 0.829] & 21526 & 21526 & 19093\\
Static EGA & Absolute dimension error & 0.799 & 0.0159 & {}[0.767, 0.829] & 21526 & 21526 & 19093\\
Static EGA & Adjusted Rand index & 0.894 & 0.0021 & {}[0.889, 0.898] & 21526 & 21526 & 19093\\
Static EGA & Variation of information & 0.147 & 0.0029 & {}[0.142, 0.153] & 21526 & 21526 & 19093\\
Static EGA & Assignment accuracy & 0.922 & 0.0015 & {}[0.919, 0.925] & 21526 & 21526 & 19093\\
\bottomrule
\end{tabular}
\end{table}
\clearpage
\begin{table}

\caption{\label{tab:grouping-t25}Louvain grouping diagnostics for the matched-method simulation with $T=25$. All diagnostics use valid network outputs. Signed dimension error is the estimated number of dimensions minus the generating number; absolute error is its magnitude. Higher adjusted Rand index and assignment accuracy, and lower variation of information, indicate closer agreement. MCSE denotes Monte Carlo standard error. The 95\% intervals resample whole datasets within each reported analysis group.}
\centering
\begin{tabular}[t]{llrrcrrr}
\toprule
Method & Diagnostic & Estimate & MCSE & 95\% interval & Datasets & Valid outputs & Correct dimensions\\
\midrule
DynEGA & Signed dimension error & 0.492 & 0.0102 & {}[0.471, 0.512] & 21600 & 21600 & 18636\\
DynEGA & Absolute dimension error & 0.494 & 0.0102 & {}[0.474, 0.515] & 21600 & 21600 & 18636\\
DynEGA & Adjusted Rand index & 0.902 & 0.0018 & {}[0.899, 0.906] & 21600 & 21600 & 18636\\
DynEGA & Variation of information & 0.137 & 0.0025 & {}[0.132, 0.143] & 21600 & 21600 & 18636\\
DynEGA & Assignment accuracy & 0.935 & 0.0012 & {}[0.933, 0.938] & 21600 & 21600 & 18636\\
\addlinespace
GraphicalVAR & Signed dimension error & 1.048 & 0.0154 & {}[1.017, 1.076] & 21579 & 21579 & 16641\\
GraphicalVAR & Absolute dimension error & 1.048 & 0.0154 & {}[1.016, 1.080] & 21579 & 21579 & 16641\\
GraphicalVAR & Adjusted Rand index & 0.859 & 0.0020 & {}[0.855, 0.863] & 21579 & 21579 & 16641\\
GraphicalVAR & Variation of information & 0.200 & 0.0028 & {}[0.195, 0.206] & 21579 & 21579 & 16641\\
GraphicalVAR & Assignment accuracy & 0.893 & 0.0015 & {}[0.890, 0.896] & 21579 & 21579 & 16641\\
\addlinespace
Static EGA & Signed dimension error & 0.576 & 0.0136 & {}[0.551, 0.603] & 21541 & 21541 & 19721\\
Static EGA & Absolute dimension error & 0.576 & 0.0136 & {}[0.550, 0.603] & 21541 & 21541 & 19721\\
Static EGA & Adjusted Rand index & 0.922 & 0.0018 & {}[0.919, 0.926] & 21541 & 21541 & 19721\\
Static EGA & Variation of information & 0.108 & 0.0025 & {}[0.103, 0.113] & 21541 & 21541 & 19721\\
Static EGA & Assignment accuracy & 0.943 & 0.0013 & {}[0.940, 0.945] & 21541 & 21541 & 19721\\
\bottomrule
\end{tabular}
\end{table}
\clearpage
\normalsize\end{landscape}
\begin{landscape}\scriptsize\setstretch{1}\setlength{\tabcolsep}{3pt}
\begin{table}

\caption{\label{tab:trajectory-t100}Louvain trajectory-summary diagnostics for the matched-method simulation with $T=100$. Rows marked correct dimensions include valid outputs that recovered the correct number of dimensions. Absolute correlation denotes best-match absolute trajectory correlation; NRMSE denotes normalized root mean squared error. Correlation may reuse a recovered-group summary for multiple latent dimensions; NRMSE uses a separate one-to-one match. All-attempt rows assign correlation 0 and NRMSE 2 when no valid network output was returned or the corresponding diagnostic was undefined. Higher correlation and lower NRMSE indicate closer alignment with the activity-scaled latent trajectories.}
\centering
\begin{tabular}[t]{p{1.00in}p{2.85in}rrcrrr}
\toprule
Method & Diagnostic & Estimate & MCSE & 95\% interval & Datasets & Valid & Correct dimension\\
\midrule
DynEGA & Absolute correlation (all attempts) & 0.728 & 0.0017 & {}[0.725, 0.731] & 21600 & 21600 & 20282\\
DynEGA & NRMSE (all attempts) & 0.642 & 0.0031 & {}[0.636, 0.647] & 21600 & 21600 & 20282\\
GraphicalVAR & Absolute correlation (all attempts) & 0.668 & 0.0020 & {}[0.664, 0.671] & 21600 & 20329 & 16784\\
GraphicalVAR & NRMSE (all attempts) & 0.753 & 0.0036 & {}[0.746, 0.760] & 21600 & 20329 & 16784\\
GIMME & Absolute correlation (all attempts) & 0.132 & 0.0019 & {}[0.129, 0.136] & 21600 & 4846 & 3795\\
\addlinespace
GIMME & NRMSE (all attempts) & 1.749 & 0.0034 & {}[1.742, 1.756] & 21600 & 4846 & 3795\\
Static EGA & Absolute correlation (all attempts) & 0.727 & 0.0017 & {}[0.723, 0.730] & 21600 & 21526 & 19093\\
Static EGA & NRMSE (all attempts) & 0.641 & 0.0031 & {}[0.635, 0.647] & 21600 & 21526 & 19093\\
DynEGA & Absolute correlation (correct dimensions) & 0.760 & 0.0015 & {}[0.757, 0.763] & 20282 & 21600 & 20282\\
DynEGA & NRMSE (correct dimensions) & 0.602 & 0.0031 & {}[0.596, 0.607] & 20282 & 21600 & 20282\\
\addlinespace
GraphicalVAR & Absolute correlation (correct dimensions) & 0.791 & 0.0014 & {}[0.789, 0.794] & 16784 & 20329 & 16784\\
GraphicalVAR & NRMSE (correct dimensions) & 0.568 & 0.0031 & {}[0.562, 0.574] & 16784 & 20329 & 16784\\
GIMME & Absolute correlation (correct dimensions) & 0.606 & 0.0034 & {}[0.600, 0.613] & 3795 & 4846 & 3795\\
GIMME & NRMSE (correct dimensions) & 0.898 & 0.0050 & {}[0.888, 0.907] & 3795 & 4846 & 3795\\
Static EGA & Absolute correlation (correct dimensions) & 0.787 & 0.0014 & {}[0.784, 0.789] & 19093 & 21526 & 19093\\
\addlinespace
Static EGA & NRMSE (correct dimensions) & 0.563 & 0.0031 & {}[0.557, 0.569] & 19093 & 21526 & 19093\\
\bottomrule
\end{tabular}
\end{table}
\clearpage
\begin{table}

\caption{\label{tab:trajectory-t25}Louvain trajectory-summary diagnostics for the matched-method simulation with $T=25$. Rows marked correct dimensions include valid outputs that recovered the correct number of dimensions. Absolute correlation denotes best-match absolute trajectory correlation; NRMSE denotes normalized root mean squared error. Correlation may reuse a recovered-group summary for multiple latent dimensions; NRMSE uses a separate one-to-one match. All-attempt rows assign correlation 0 and NRMSE 2 when no valid network output was returned or the corresponding diagnostic was undefined. Higher correlation and lower NRMSE indicate closer alignment with the activity-scaled latent trajectories.}
\centering
\begin{tabular}[t]{p{1.00in}p{2.85in}rrcrrr}
\toprule
Method & Diagnostic & Estimate & MCSE & 95\% interval & Datasets & Valid & Correct dimension\\
\midrule
DynEGA & Absolute correlation (all attempts) & 0.800 & 0.0012 & {}[0.797, 0.802] & 21600 & 21600 & 18636\\
DynEGA & NRMSE (all attempts) & 0.593 & 0.0023 & {}[0.589, 0.598] & 21600 & 21600 & 18636\\
GraphicalVAR & Absolute correlation (all attempts) & 0.799 & 0.0012 & {}[0.796, 0.801] & 21600 & 21579 & 16641\\
GraphicalVAR & NRMSE (all attempts) & 0.587 & 0.0023 & {}[0.582, 0.591] & 21600 & 21579 & 16641\\
GIMME & Absolute correlation (all attempts) & 0.000 & 0.0000 & {}[0.000, 0.000] & 21600 & 0 & 0\\
\addlinespace
GIMME & NRMSE (all attempts) & 2.000 & 0.0000 & {}[2.000, 2.000] & 21600 & 0 & 0\\
Static EGA & Absolute correlation (all attempts) & 0.801 & 0.0012 & {}[0.799, 0.803] & 21600 & 21541 & 19721\\
Static EGA & NRMSE (all attempts) & 0.595 & 0.0024 & {}[0.591, 0.600] & 21600 & 21541 & 19721\\
DynEGA & Absolute correlation (correct dimensions) & 0.848 & 0.0009 & {}[0.846, 0.850] & 18636 & 21600 & 18636\\
DynEGA & NRMSE (correct dimensions) & 0.521 & 0.0023 & {}[0.517, 0.526] & 18636 & 21600 & 18636\\
\addlinespace
GraphicalVAR & Absolute correlation (correct dimensions) & 0.871 & 0.0008 & {}[0.870, 0.873] & 16641 & 21579 & 16641\\
GraphicalVAR & NRMSE (correct dimensions) & 0.466 & 0.0022 & {}[0.461, 0.470] & 16641 & 21579 & 16641\\
Static EGA & Absolute correlation (correct dimensions) & 0.836 & 0.0009 & {}[0.834, 0.838] & 19721 & 21541 & 19721\\
Static EGA & NRMSE (correct dimensions) & 0.548 & 0.0023 & {}[0.544, 0.553] & 19721 & 21541 & 19721\\
\bottomrule
\end{tabular}
\end{table}
\clearpage
\normalsize\end{landscape}

\newpage

\subsection{C.3 Sensitivity Analyses and Realized
Amplification}\label{c.3-sensitivity-analyses-and-realized-amplification}

\begin{figure}[H]

\centering{

\includegraphics[width=0.95\linewidth,height=\textheight,keepaspectratio]{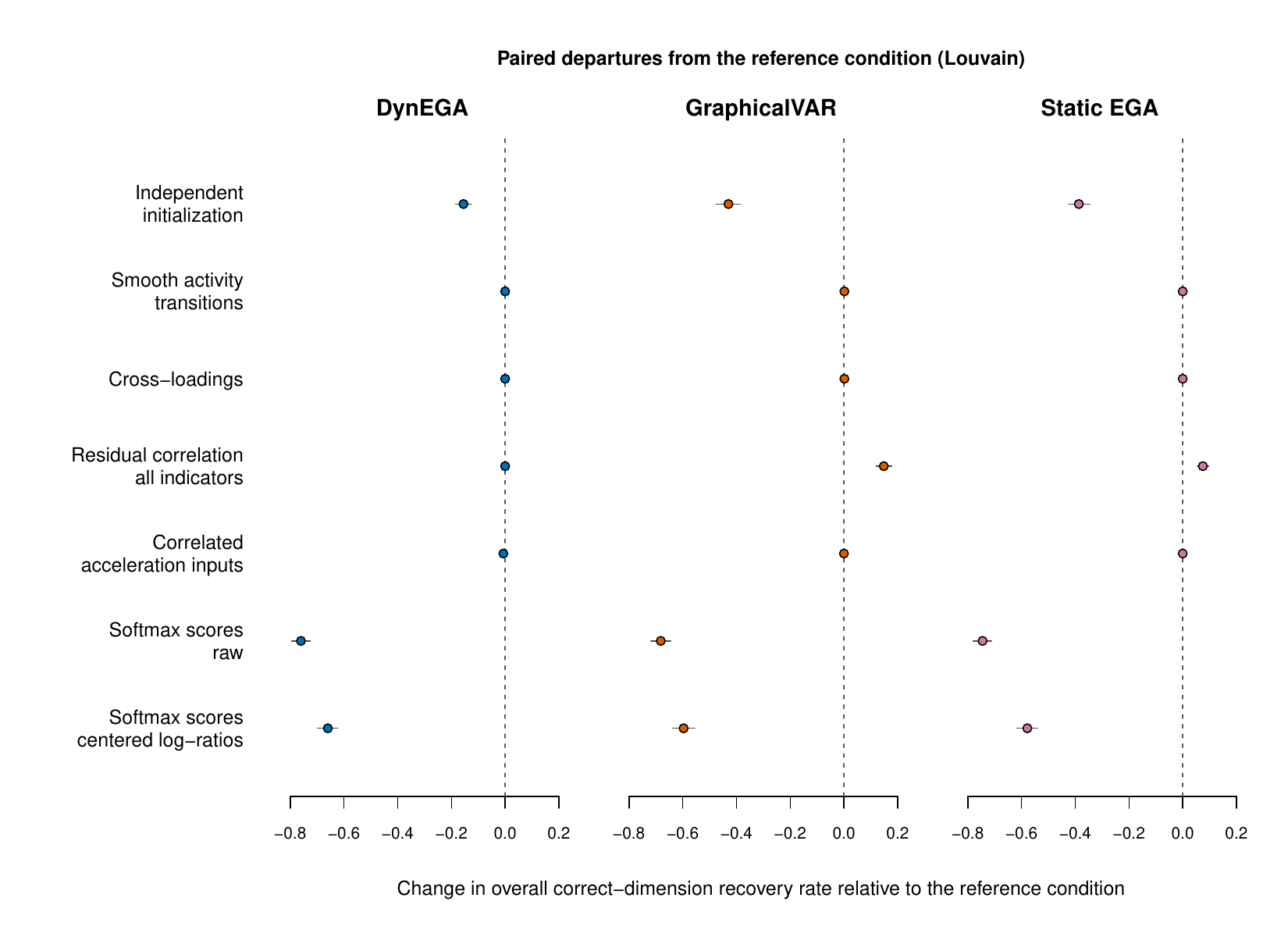}

}

\caption{\label{fig-factor-effects}Paired change in overall
correct-dimension recovery under each one-at-a-time sensitivity
condition. Negative values indicate lower recovery than in the reference
condition. Error bars are 95\% bootstrap intervals from resampling whole
datasets.}

\end{figure}%

Table B2 gives the dynamic parameters used in these scenarios. Table C5
summarizes the amplitudes generated in the amplifying families E and F.
Table C6 describes deterministic growth or decay for each distinct
parameter pair. Together, these diagnostics distinguish the specified
dynamic regime from the amplitudes observed in finite simulated records.

\begingroup\scriptsize\setlength{\tabcolsep}{2pt}

\begin{longtable}[t]{crrrp{0.85in}p{0.95in}p{0.85in}p{0.85in}}
\caption{Realized amplitude diagnostics for the amplifying families E and F (Table B1) with $N=50$, four indicators per dimension, loading 0.60, residual standard deviation 0.25, and 25 replications. The final-tenth energy fraction is the proportion of squared activity-scaled latent signal occurring in the last 10\% of observations, averaged across clips within a panel and then summarized across replications.}\\
\toprule
Code & Dimensions & $T$ & $H$ & Median max $|\tilde{x}|$ & 95th percentile max $|\tilde{x}|$ & Median observed IQR & Final-tenth energy fraction\\
\midrule
\endfirsthead
\toprule
Code & Dimensions & $T$ & $H$ & Median max $|\tilde{x}|$ & 95th percentile max $|\tilde{x}|$ & Median observed IQR & Final-tenth energy fraction\\
\midrule
\endhead
E & 2 & 100 & 39.60 & 55.9 & 61.5 & 3.47 & 0.316\\
E & 2 & 300 & 74.75 & 329.4 & 376.9 & 7.46 & 0.338\\
E & 3 & 100 & 39.60 & 62.6 & 73.0 & 1.90 & 0.368\\
E & 3 & 300 & 74.75 & 430.3 & 478.6 & 4.58 & 0.428\\
E & 4 & 100 & 39.60 & 85.6 & 102.7 & 1.57 & 0.404\\
\addlinespace
E & 4 & 300 & 74.75 & 994.9 & 1092.7 & 4.02 & 0.498\\
F & 2 & 100 & 39.60 & 69.5 & 81.9 & 4.71 & 0.319\\
F & 2 & 300 & 74.75 & 388.1 & 491.3 & 9.31 & 0.324\\
F & 3 & 100 & 39.60 & 77.7 & 87.5 & 2.33 & 0.367\\
F & 3 & 300 & 74.75 & 518.5 & 608.5 & 5.74 & 0.448\\
\addlinespace
F & 4 & 100 & 39.60 & 105.2 & 122.2 & 1.92 & 0.402\\
F & 4 & 300 & 74.75 & 1201.6 & 1350.2 & 4.97 & 0.517\\
\bottomrule
\end{longtable}
\endgroup

\begin{landscape}\footnotesize\setlength{\tabcolsep}{3pt}

\begin{longtable}[t]{lrrrrrrrrp{1.30in}}
\caption{Deterministic stability diagnostics for the unique dynamic-parameter pairs used across scenario factors. Table B2 maps these pairs to scenarios and dimensions. The spectral radius $\rho(\Phi)$ is the largest absolute eigenvalue of the transition matrix; values below one indicate decay. The modal exponential-rate factors $g(H)=\exp\{H\max\operatorname{Re}(r)\}$ describe modal rates over the observed horizon rather than total finite-horizon state amplification. $H[100]=39.6$ and $H[300]=74.75$ under the stated record designs.}\\
\toprule
Scenario codes & eta & zeta & Discriminant & Max Re(root) & rho(Phi[.4]) & rho(Phi[.25]) & g(H[100]) & g(H[300]) & Regime\\
\midrule
\endfirsthead
\toprule
Scenario codes & eta & zeta & Discriminant & Max Re(root) & rho(Phi[.4]) & rho(Phi[.25]) & g(H[100]) & g(H[300]) & Regime\\
\midrule
\endhead
A, B, C, D & -1.10 & -0.30 & -4.3100 & -0.150 & 0.9418 & 0.9632 & 0.0026 & 0.0000 & Stable oscillatory\\
A, B, C, D & -0.85 & -0.65 & -2.9775 & -0.325 & 0.8781 & 0.9220 & 0.0000 & 0.0000 & Stable oscillatory\\
A, B, C, D & -0.55 & -1.00 & -1.2000 & -0.500 & 0.8187 & 0.8825 & 0.0000 & 0.0000 & Stable oscillatory\\
A, B, C, D & -0.30 & -1.70 & 1.6900 & -0.200 & 0.9231 & 0.9512 & 0.0004 & 0.0000 & Stable nonoscillatory\\
E, F & -0.20 & 0.14 & -0.7804 & 0.070 & 1.0284 & 1.0177 & 15.9906 & 187.2604 & Finite-horizon amplifying oscillatory\\
\addlinespace
E, F & -0.15 & 0.12 & -0.5856 & 0.060 & 1.0243 & 1.0151 & 10.7618 & 88.6770 & Finite-horizon amplifying oscillatory\\
E, F & -0.10 & 0.10 & -0.3900 & 0.050 & 1.0202 & 1.0126 & 7.2427 & 41.9929 & Finite-horizon amplifying oscillatory\\
E, F & -0.05 & 0.08 & -0.1936 & 0.040 & 1.0161 & 1.0101 & 4.8744 & 19.8857 & Finite-horizon amplifying oscillatory\\
\bottomrule
\end{longtable}
\end{landscape}

\newpage

\subsection{C.4 Extended Factorial
Simulation}\label{c.4-extended-factorial-simulation}

The extended factorial simulation evaluated DynEGA on 518,400 datasets.
It recovered the correct dimension in 93.7\% of those datasets and
assigned 95.2\% of indicators correctly. Stronger loadings and smaller
residual error improved recovery. The design, including the \(T=300\)
change in observation interval, is given in Appendix B. Runtime
accounting for this design is in Section C.5.

\subsection{C.5 Runtime and Feasibility
Accounting}\label{c.5-runtime-and-feasibility-accounting}

\begin{figure}[H]

\centering{

\includegraphics[width=0.95\linewidth,height=\textheight,keepaspectratio]{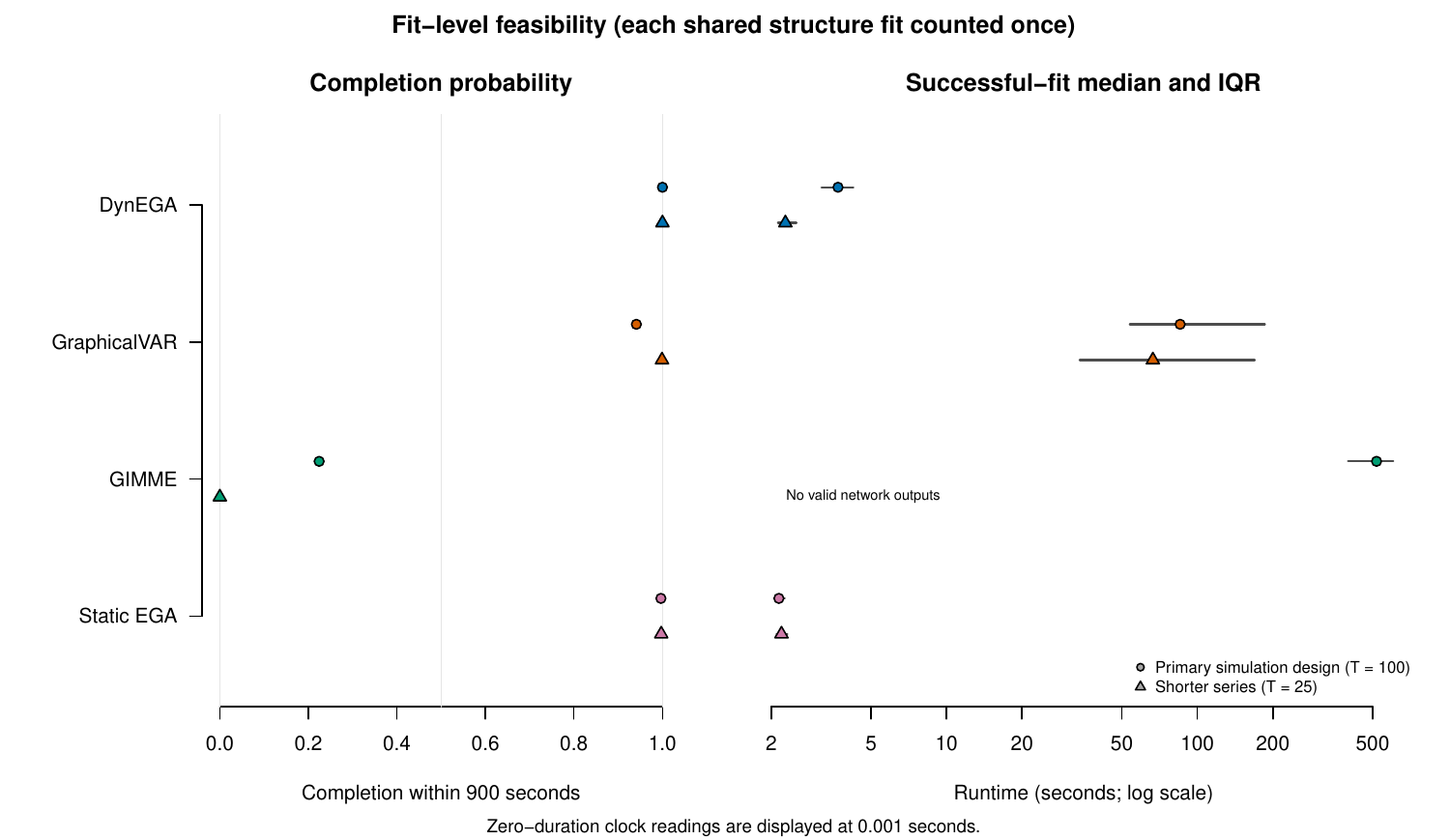}

}

\caption{\label{fig-runtime-feasibility}Runtime among valid network
outputs in the primary \(T=100\) and shorter \(T=25\) designs. Points
show medians and intervals show interquartile ranges on a log10 scale.
Counts use unique fitted structures.}

\end{figure}%

\begin{landscape}\footnotesize\setlength{\tabcolsep}{3pt}

\begin{longtable}[t]{lclrrrrcc}
\caption{Runtime and feasibility per method fit in the $T=100$ primary and $T=25$ shorter-series matched-method designs. Valid-output intervals resample whole simulated datasets. Runtime summaries condition on valid network outputs and include structure estimation, network extraction, and community detection. Each fitted structure is reused for Louvain, Leiden, and Walktrap.}\\
\toprule
Analysis & $T$ & Method & Attempts & Valid outputs & Timeouts & Errors & Valid-output rate & Runtime median [IQR], s\\
\midrule
\endfirsthead
\toprule
Analysis & $T$ & Method & Attempts & Valid outputs & Timeouts & Errors & Valid-output rate & Runtime median [IQR], s\\
\midrule
\endhead
Matched methods & $T=100$ & DynEGA & 21600 & 21600 & 0 & 0 & 1.000 [1.000, 1.000] & 3.69 [3.18, 4.25]\\
Matched methods & $T=100$ & GIMME & 21600 & 4846 & 14477 & 2277 & 0.224 [0.219, 0.230] & 518.16 [398.98, 603.71]\\
Matched methods & $T=100$ & GraphicalVAR & 21600 & 20329 & 1271 & 0 & 0.941 [0.938, 0.944] & 85.41 [53.82, 185.91]\\
Matched methods & $T=100$ & Static EGA & 21600 & 21526 & 0 & 74 & 0.997 [0.996, 0.997] & 2.15 [2.05, 2.26]\\
Matched methods & $T=25$ & DynEGA & 21600 & 21600 & 0 & 0 & 1.000 [1.000, 1.000] & 2.28 [2.13, 2.52]\\
\addlinespace
Matched methods & $T=25$ & GIMME & 21600 & 0 & 0 & 21600 & 0.000 [0.000, 0.000] & Not available\\
Matched methods & $T=25$ & GraphicalVAR & 21600 & 21579 & 21 & 0 & 0.999 [0.999, 0.999] & 66.49 [33.96, 169.39]\\
Matched methods & $T=25$ & Static EGA & 21600 & 21541 & 0 & 59 & 0.997 [0.997, 0.998] & 2.20 [2.10, 2.32]\\
\bottomrule
\end{longtable}
\end{landscape}

\newpage

\subsection{C.6 Community-Detection
Sensitivity}\label{c.6-community-detection-sensitivity}

\begin{figure}[H]

\centering{

\includegraphics[width=0.98\linewidth,height=\textheight,keepaspectratio]{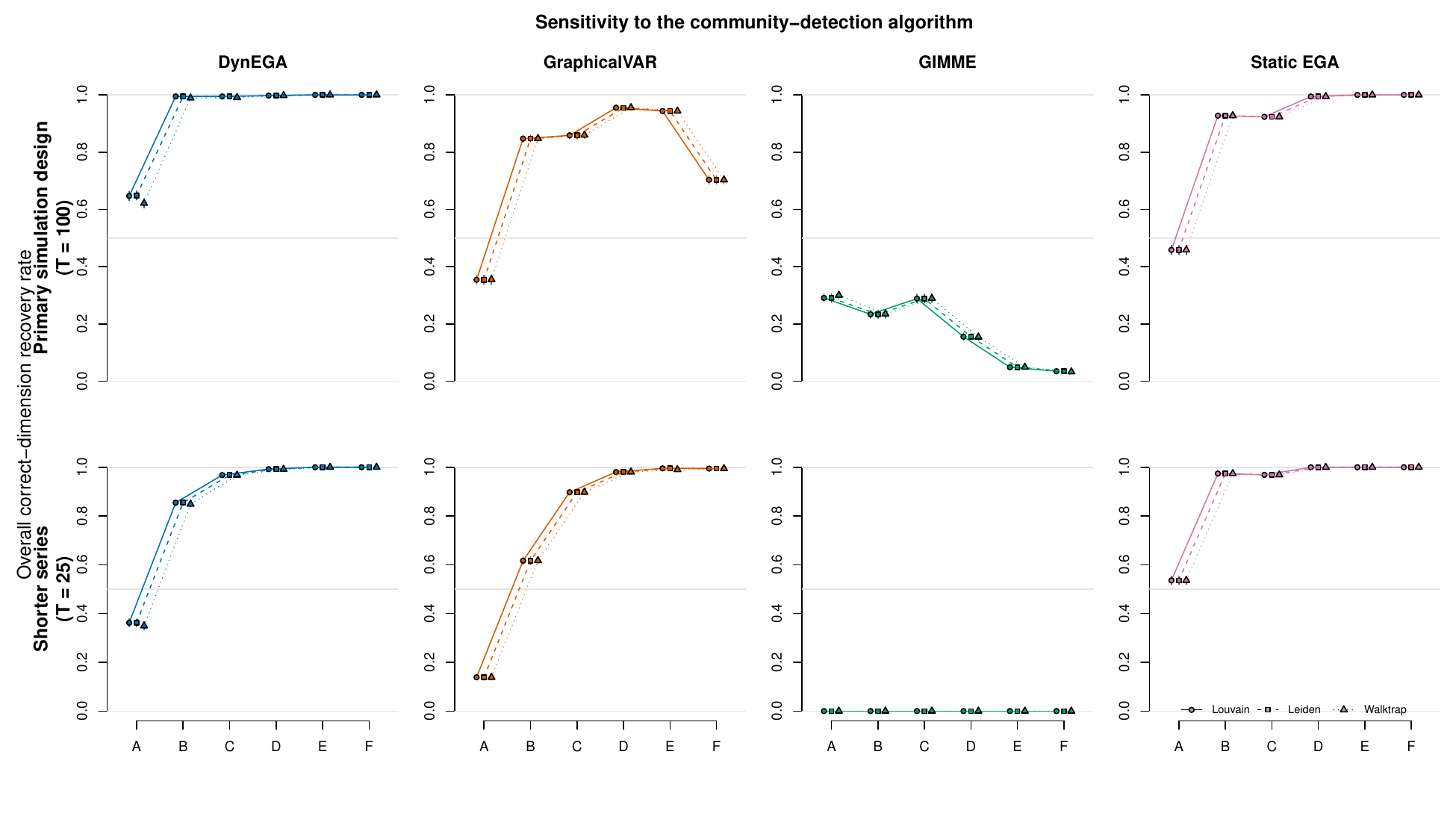}

}

\caption{\label{fig-community-sensitivity}Overall correct-dimension
recovery rates for Louvain, Leiden, and Walktrap in the \(T=100\)
primary and \(T=25\) shorter-series matched-method designs. Each fitted
network is reused across all three algorithms, so differences isolate
the community-detection step.}

\end{figure}

\begin{figure}[H]

\centering{

\includegraphics[width=0.98\linewidth,height=\textheight,keepaspectratio]{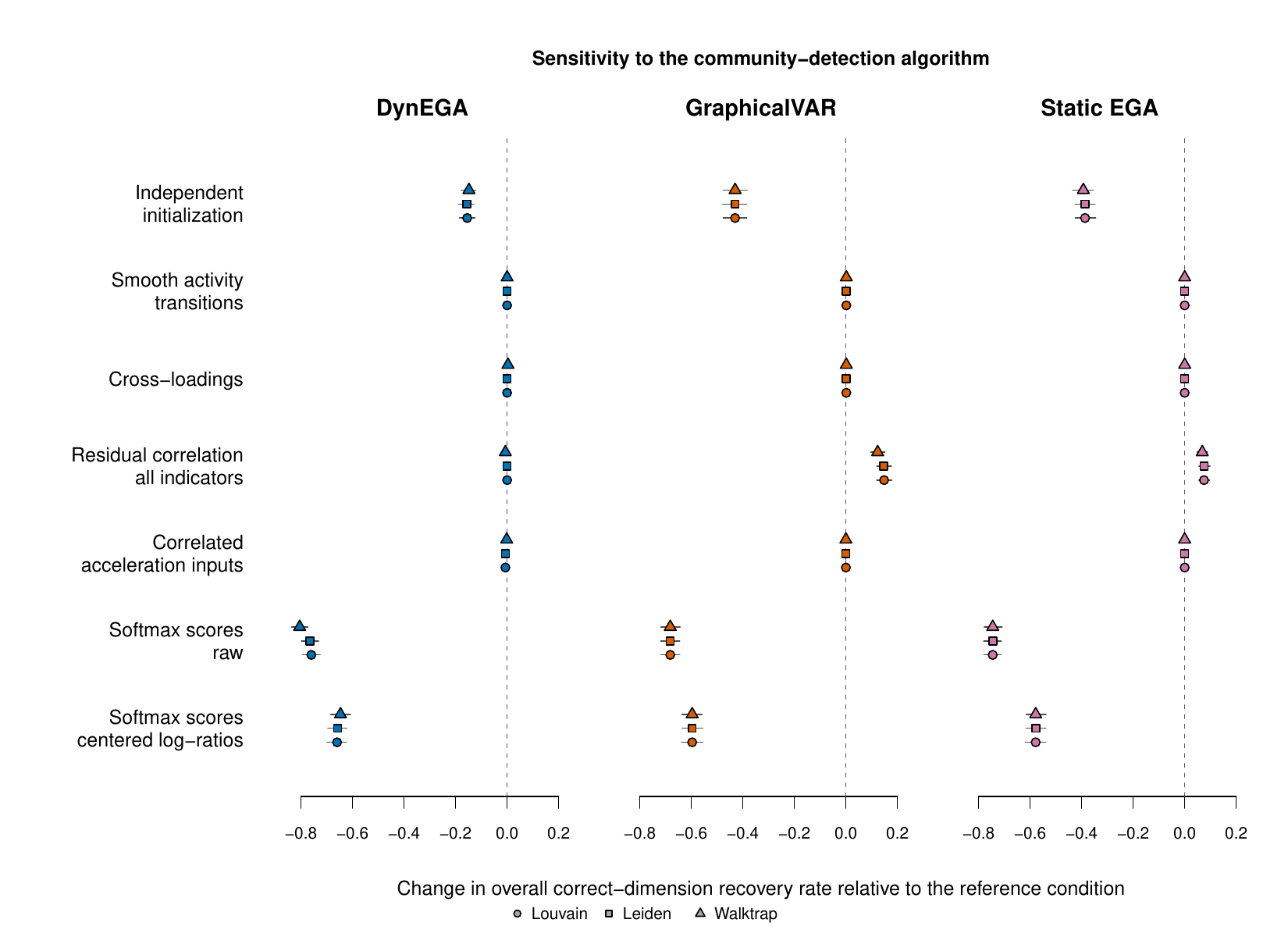}

}

\caption{\label{fig-challenge-community-sensitivity}Sensitivity of the
paired one-at-a-time effects to Louvain, Leiden, and Walktrap. Values
are changes in overall correct-dimension recovery rate relative to the
reference condition.}

\end{figure}

Louvain, Leiden, and Walktrap were applied to each retained fitted
graph. Their results were closely aligned in the matched-method
simulation designs (Figure~\ref{fig-community-sensitivity}) and led to
the same interpretation of the one-at-a-time sensitivity conditions
(Figure~\ref{fig-challenge-community-sensitivity}). These comparisons
reuse fitted structures and therefore assess the community-detection
step rather than estimator variability.

\end{document}